\documentclass[twocolumn]{aastex63}
\usepackage{epsfig}
\usepackage{natbib}
\usepackage{amsmath}
\usepackage{hyperref}
\usepackage{lineno}
\usepackage{xcolor}

\newcommand{\Msun}{M_{\odot}}

\newcommand{\kms}{\mbox{${\rm km~s^{-1}}$}}
\newcommand{\kmsMpc}{\mbox{km s$^{-1}$ Mpc$^{-1}$}}

\newcommand{\p}{^\prime}
\newcommand{\Pc}{\mathcal{P}}
\newcommand{\Dc}{\mathcal{D}}
\newcommand{\Lc}{\mathcal{L}}
\newcommand{\hi}{H{\sc\,i} }

\begin{document}

\title{Cosmicflows-4}

\author{R. Brent Tully}
\affil{Institute for Astronomy, University of Hawaii, 2680 Woodlawn Drive, Honolulu, HI 96822, USA}
\author{Ehsan Kourkchi}
\affil{Institute for Astronomy, University of Hawaii, 2680 Woodlawn Drive, Honolulu, HI 96822, USA}
\author{H\'el\`ene M. Courtois}
\affil{University of Lyon, UCB Lyon 1, CNRS/IN2P3, IUF, IP2I Lyon, France}
\author{Gagandeep S. Anand}
\affil{Space Telescope Science Institute, 3700 San Martin Drive, Baltimore, MD 21218, USA}
\author{John P. Blakeslee}
\affil{Gemini Observatory \& NSF's NOIRLab, 950 N. Cherry Ave., Tucson, AZ 85719, USA}
\author{Dillon Brout}
\affil{Center for Astrophysics, Harvard \& Smithsonian, 60 Garden St., Cambridge, MA 02138, USA}
\author{Thomas de Jaeger}
\affil{Institute for Astronomy, University of Hawaii, 2680 Woodlawn Drive, Honolulu, HI 96822, USA}
\author{Alexandra Dupuy}
\affil{Korea Institute for Advanced Study, 85, Hoegi-ro, Dongdaemun-gu, Seoul 02455, Republic of Korea}
\author{Daniel Guinet}
\affil{University of Lyon, UCB Lyon 1, CNRS/IN2P3, IUF, IP2I Lyon, France}
\author{Cullan Howlett}
\affil{School of Mathematics and Physics, The University of Queensland, Brisbane, QLD 4072, Australia.}
\author{Joseph B. Jensen}
\affil{Department of Physics, Utah Valley University, 800 W. University Parkway, Orem, UT 84058, USA}
\author{Daniel Pomar\`ede}
\affil{Institut de Recherche sur les Lois Fondamentales de l'Univers, CEA Universit\'e Paris-Saclay, 91191 Gif-sur-Yvette, France}
\author{Luca Rizzi}
\affil{W.M. Keck Observatory, 65-1120 Mamalahoa Highway, Kamuela, HI 96743, USA}
\author{David Rubin}
\affil{Department of Physics \& Astronomy, University of Hawaii at Manoa, Honolulu, HI 96822, USA}
\author{Khaled Said}
\affil{School of Mathematics and Physics, The University of Queensland, Brisbane, QLD 4072, Australia.}
\author{Daniel Scolnic}
\affil{Department of Physics, Duke University, Durham, NC 27708, USA}
\author{Benjamin E. Stahl}
\affil{Department of Astronomy, University of California, CA 94720-3411, USA}

\begin{abstract}
With {\it Cosmicflows-4}, distances are compiled for 55,877 galaxies gathered into 38,065 groups.  Eight methodologies are employed, with the largest numbers coming from the correlations between the photometric and kinematic properties of spiral galaxies (TF) and elliptical galaxies (FP).  Supernovae that arise from degenerate progenitors (type Ia Sne) are an important overlapping component.  Smaller contributions come from distance estimates from the surface brightness fluctuations of elliptical galaxies and the luminosities and expansion rates of core collapse supernovae (SNII).  Cepheid period-luminosity relation and tip of the red giant branch observations founded on local stellar parallax measurements along with the geometric maser distance to NGC\,4258 provide the absolute scaling of distances.  The assembly of galaxies into groups is an important feature of the study in facilitating overlaps between methodologies.  Merging between multiple contributions within a methodology and between methodologies is carried out with Bayesian Markov chain Monte Carlo procedures.  The final assembly of distances is compatible with a value of the Hubble constant of $H_0=74.6$~\kmsMpc\ with the small statistical error of $\pm0.8$~\kmsMpc\ but a large potential systematic error of $\sim 3$~\kmsMpc. Peculiar velocities can be inferred from the measured distances.  The interpretation of the field of peculiar velocities is complex because of large errors on individual components and invites analyses beyond the scope of this study.

\end{abstract}



\bigskip
\section{Introduction}
\label{sec:intro}

{\it Cosmicflows} is a program to compile galaxy distances and parse observed velocities into components due to the expansion of the universe and residuals due to gravitational interactions.  Our fundamental interest is to derive inferences regarding the large-scale structure of the universe from galaxy test particle peculiar motions.  This fourth release of the program follows those of \citet{2008ApJ...676..184T, 2013AJ....146...86T, 2016AJ....152...50T}.

Contributions to the {\it Cosmicflows} program have come from work within our collaboration and from the literature.  We consider methodologies that have been tested and have physical bases that are reasonably well understood.  It is as great a consideration, though, that there be large overlaps between contributions.  A sample with distances to only a few objects cannot confidently be meshed within a common scale so is not very useful.

We derive distances in significant numbers mainly from seven methodologies.  By far, the largest quantitative contributions are given by the fundamental plane (FP) correlation between the luminosity, surface brightness, and central velocity dispersion of early-type galaxies \citep{1987ApJ...313...42D, 1987ApJ...313...59D} and the luminosity-rotation rate relation for spiral galaxies \citep{1977A&A....54..661T} (TF or Tully-Fisher relation (TFR)).  The individual errors in these cases are substantial ($20-25\%$) but the objects are widely dispersed, providing a dense network of distance information across the sky extending to $\sim0.05c$ and in the celestial and galactic north to $z=0.1$, the upper cutoff of our compilation.

Three other methods probe substantial distances with greater accuracy but their contributions remain small.  Type~Ia supernovae (SNIa) \citep{1993ApJ...413L.105P} provide distances with an accuracy of $\sim7\%$ out to $0.1c$.  Type-II supernovae (SNII) \citep{2002ApJ...566L..63H} provide distances with $\sim15\%$ accuracy to similar distances.  Surface brightness fluctuations (SBF) monitoring the degree of resolution of the old stellar populations of elliptical galaxies \citep{1988AJ.....96..807T} can provide distances with $\sim5\%$ accuracy to targets within $\sim0.03c$.

While comparisons between these five methods can be set on a common relative scale, it remains to provide them an absolute calibration.  Two methods provide a bridge: those provided by the Cepheid period-luminosity relation (CPLR) \citep{1912HarCi.173....1L} and the constancy of stellar luminosities at the tip of the red giant branch (TRGB) \citep{1993ApJ...417..553L}.  These methods provide accurate distances ($\sim5\%$) but are restricted to less than $\sim 20$~Mpc. 

There has been considerable effort to establish the absolute scale of the CPLR and TRGB procedures through geometrically based observations.  Parallax distances can be established to Cepheids within our own galaxy \citep{2007AJ....133.1810B} and parallax distances to RR Lyrae and horizontal branch (HB) stars can establish the TRGB scale \citep{2007ApJ...661..815R}.  It is anticipated that observations with GAIA \citep{2018ASPC..514...89C, 2019PASA...36....1M} will provide robust direct Cepheid and TRGB calibrations in the near future.  Meanwhile, important links to an absolute scale are provided by detached eclipsing binaries in the Large Magellanic Cloud (LMC) \citep{2019Natur.567..200P} and the maser system in the nuclear region of the galaxy NGC~4258 \citep{2019ApJ...886L..27R}.

The main contributions in the first version of {\it Cosmicflows} \citep{2008ApJ...676..184T} were based on the TFR with optical photometry obtained object by object and analog neutral hydrogen (\hi) linewidths.  The catalog contained  distances to 1,791 galaxies constrained to the limit 3,000~\kms.

{\it Cosmicflows-2} \citep{2013AJ....146...86T} was expanded to include a much larger volume, peaking in numbers at 5,000~\kms\ with a tail extending to $\sim15,000$~\kms.  Most of the contributions came from the TFR, with as a major revision the employment of a rigorous algorithm in the reduction of digital \hi spectra \citep{2009AJ....138.1938C, 2011MNRAS.414.2005C}.  Likewise, the photometric analysis was more rigorously defined \citep{2011MNRAS.415.1935C, 2012AJ....144..133S}.  The catalog then grew to include 8,188 galaxies.

The major addition to {\it Cosmicflows-3} \citep{2016AJ....152...50T} was FP distance measures from the 6-degree Field Galaxy Survey (6dFGSv) \citep{2014MNRAS.445.2677S}.  This sample is entirely confined to the celestial south and abruptly cuts off at 16,000~\kms.  A secondary addition came from the TFR method with infrared photometry provided by the Spitzer Space Telescope \citep{2014MNRAS.444..527S}.  {\it Cosmicflows-3} provided distances for 17,699 galaxies.  Coverage within $\sim8,000$~\kms\ was reasonably balanced around the sky but at 8,000-16,000~\kms\ it strongly favored the southern hemisphere.  The infrared TFR contribution was confined to within $\sim6,000$~\kms\ but notably extended coverage to low galactic latitudes, shrinking the coverage gap between galactic hemispheres.

Here, with {\it Cosmicflows-4} a most important addition is a much-extended TFR sample of 10,000 galaxies drawing in particular on kinematic information from ALFALFA, the Arecibo Legacy Fast ALFA survey of the high galactic latitude sky in the decl. range $0 - 38$~degrees \citep{2011AJ....142..170H, 2018ApJ...861...49H}.  Photometry is provided by SDSS, the Sloan Digital Sky Survey \citep{2000AJ....120.1579Y} and WISE, the Wide-field Infrared Explorer \citep{2010AJ....140.1868W}.  This component of {\it Cosmicflows-4} substantially redresses the imbalance favoring the southern sky of the previous catalog.

SDSS also provides the source material for a second even larger addition to the current catalog.  SDSS photometry and spectroscopy are combined to provide FP distances to 34,000 galaxies out to 30,000~\kms\ in the quadrant of the sky that is celestial north and galactic north.  As a consequence, while {\it Cosmicflows-3} tilted toward coverage of the celestial south, now {\it Cosmicflows-4} greatly expands our knowledge of the north.

With the astronomical community's overriding interest in precision distance measurements in order to secure the value of the Hubble constant, there are understandable arguments for a maximally homogeneous approach \citep{2016ApJ...826...56R}. The {\it Cosmicflows} assembly is heterogeneous.  It is to be appreciated that the primary interest of this program is the mapping of {\it deviations} from cosmic expansion, requiring coherence of distance measurements but not an absolute scaling.  Nonetheless, the reasonable establishment of a zero-point is not our most difficult task.  Our heterogeneous approach has virtues.  Results from separate methodologies can be compared by sectors of the sky or distance, potentially revealing systematics.  Different contributions favor ancient populations or young, members of clusters or the field.  Some are better probes of low galactic latitudes.  Our samples are heterogeneous but not indiscriminate.

Coincidences of distance measurements by different methodologies to members of a common group enables the stitching of samples into a coherent ensemble.  Our discussion will turn first to the important matter of the definition of groups in \S\ref{sec:groups}.
Subsequent sections will focus on each of the seven methodologies that provide most of our distances.  We begin with the numerically dominant TFR (\S\ref{sec:TFR}) and FP (\S\ref{sec:FP}) components, benefiting from overlaps between large samples to establish coherence in a core compilation. We then focus on the SBF (\S\ref{sec:sbf}) and SN~Ia (\S\ref{sec:sn}) contributions that are modest in number but that impose demanding constraints. There is a brief discussion of SN~II (\S\ref{sec:snii}) that at this point makes a relatively small contribution.  This entire edifice is then be linked to foundational TRGB (\S\ref{sec:trgb}) and CPLR (\S\ref{sec:cplr}) information, these in turn grounded by geometrical maser\footnote{Studies of nuclear maser systems provide an eighth methodology and one that gives independent absolute distance estimates but these are only available for six galaxies discussed in \S\ref{sec:maser}.} , eclipsing binary, and parallax observations.  The integration of methodologies is discussed in \S \ref{sec:all}.  The data products and a brief description of properties are discussed in Sections \ref{sec:catalog} and  \ref{sec:properties}.  Then, \S \ref{sec:summary} provides a summary.


\section{Galaxy Groups}
\label{sec:groups}

Galaxies tend to lie in groups, large and small.  If associations are made correctly, then all distance measures to galaxies in a group should be the same within uncertainties.  The composition of galaxy groups, then, is of major importance for our study for at least three reasons.  First, averaging over the properties of a group reduces errors.  With weighted averaging of distances, uncertainties can be brought down from single case 20-25\% values (depending on the methodology) to statistical uncertainties of a few percent with some rich clusters.  The gains apply to velocities as well.  Velocity averaging can encompass {\it all} known group members, not just those with measured distances.  Second, it is particularly important in the modeling of galaxy flows to accurately locate the rich clusters, where distance measures from at least the FP and SBF targets congregate.  Rich clusters tend to lie at focal points of galaxy streams.

Third, and perhaps most importantly, it is through the groups that we most effectively match the zero-point scaling of the diverse samples.  Given the potential for systematics within samples, across sectors of the sky, and with distance, the more overlap the better.  While there can be some overlap at the level of individual galaxies, by far most of our overlaps occur at the level of groups and clusters.  A corollary benefit is the ability to weed out egregiously bad data while comparing distances to objects in common, 

Galaxy groups come in a wide range of scales.  We want to benefit from the advantages of grouping over the full range down to the instances of pairs.  Typical friends-of-friends and related group algorithms do not scale physically over the three decades of mass of interest ($10^{12} - 10^{15}~\Msun$).  Appreciating the importance of the matter, we initiated studies that resulted in three papers.  In the first \citep{2015AJ....149...54T}, we  provided detailed account of eight well-studied groups/clusters ranging from the Local Group to the Coma Cluster.  It was possible in these eight clean cases to isolate an observable proxy for the virial radius of collapsed halos; the radius of the second turnaround, $R_{2t}$ (related to the {\it splashback} radius \citep{2014JCAP...11..019A}).  This radius is found to scale, as theory predicts, as $R_{2t} \propto M^{1/3}$ and $R_{2t} \propto \sigma_p$ for halos (groups, clusters) with mass $M$ and velocity dispersion $\sigma_p$.  This first paper establishes the coefficients of the scaling relationships.

In the second paper \citep{2015AJ....149..171T}, the scaling relationships were applied to build a group catalog involving 43,000 galaxies in the Two Micron All Sky Survey (2MASS) redshift survey essentially complete over the sky at $\vert b \vert > 5^{\circ}$ for galaxies brighter than $K_s = 11.75$ \citep{2012ApJS..199...26H}.  The 2MASS survey, given its sensitivity to old stars, provides a good representation of the mass distribution within the volume extending to 15,000~\kms\ which is our principal concern.  Relative distances are based on redshifts.  We favor the use of this group catalog at systemic velocities greater than $\sim 3000$~\kms.  Were we to consider an alternative, we would use the similarly physically motivated catalog by \citet{2017MNRAS.470.2982L}.

Nearer than $\sim 3000$~\kms\ confusion arising from peculiar velocities is severe and we have knowledge of a profusion of low surface brightness galaxies that failed to be entered into the 2MASS catalog.  Hence, for the nearby volume we turn to the group catalog assembled in the third paper \citep{2017ApJ...843...16K}, based on a heterogeneous collection of all 15,000 galaxies with known velocities within 3,500~\kms.  The groups are constituted based on the same scaling relations.  The availability of distance information from {\it Cosmicflows-3} is tremendously helpful in resolving confusion issues and evaluating masses, and hence, scaling parameters.

With both the near and far catalogs, the groups are roughly bounded by the radius of second turnaround.  Hence, they represent collapsed halos.  As a naming convention, we identify a group by the Principal Galaxies Catalog number \citep{1996A&A...311...12P} of the dominant member, which we call 1PGC.  There can be inconsistencies between the two catalogs within 3,500~\kms.  In such cases, we favor the specifications by \citet{2017ApJ...843...16K}.

These group catalogs provide an excellent description of clustering within $\sim15,000$~\kms, the useful range of the 2MASS $K_s=11.75$ redshift survey.  However, the SDSS-based FP sample extends to $0.1c$.  This SDSS FP sample is a subcomponent of the \citet{2014A&A...566A...1T,2017A&A...602A.100T} SDSS group catalogs.  The supernova samples also extend to $z=0.1$, well beyond the range of the \citet{2015AJ....149..171T} groups.  If a group affiliation is unavailable for a galaxy within the groups described above then we opt for memberships in the 2017 Tempel et al. catalog.  The relevant galaxies have PGC identifications.  We take as the 1PGC name for a Tempel et al. group the PGC number of the brightest member within our catalog.


\section{Luminosity-Linewidth Distances to Spiral Galaxies}
\label{sec:TFR}

Distances derived from the TFR are an extremely important component of the {\it Cosmicflows} program.  They are numerous and the most widely distributed.  Spiral galaxies are found in all environments, providing links with other methodologies in groups and sparse but invaluable coverage in voids.  


\subsection{The Baryonic TFR}

Conventionally in the past, TFR samples have been acquired by individual targeting of selected candidates for both the photometric (imaging) and kinematic (linewidth) required components.  Nowadays, wide-field optical, infrared, and \hi radio surveys provide access to much larger samples.  Specifically, here, we make use of serendipitous SDSS DR12 $u,g,r,i,z$ optical imaging \citep{2015ApJS..219...12A}, WISE $W1$ and $W2$ infrared imaging \citep{2010AJ....140.1868W, 2011ApJ...731...53M}, and Arecibo Legacy Fast ALFA Survey (ALFALFA) neutral hydrogen spectral detections \citep{2018ApJ...861...49H}, supplemented in the radio with pointed observations with the Green Bank Telescope and Parkes Telescope \citep{2021A&A...646A.113D}.  

In a series of three papers, we explored the properties of the optical and infrared photometric material particularly pertaining to issues of extinction \citep{2019ApJ...884...82K}, then provided optical and infrared calibrations of the TFR based on $\sim 600$ galaxies in 20 clusters \citep{2020ApJ...896....3K}, and then used the calibrations to derived TFR distances for $\sim 10,000$ galaxies \citep{2020ApJ...902..145K}.  It was subsequently revealed that the distances in the latter publication are affected by a bias: a trend in Hubble parameter values, $H_i=f_icz_i/d_i$, as functions of apparent magnitude.\footnote{The cosmological model parameter $f_i$ is defined in connection with Eq.~\ref{eq:logh}.} The bias strongly affects distance estimates to intrinsically fainter galaxies within 4,000~\kms\ and arises from faint end curvature in the TFR.  Distances to galaxies with velocities greater than 4,000~\kms\ are mildly affected.

Much, and even a majority, of baryonic mass in faint galaxies is in the form of interstellar gas.  It has been noted that adding this constituent to the stellar component represented by optical or infrared light, formulating the baryonic Tully-Fisher relation (BTFR), effectively linearizes the relation between the logarithms of baryonic mass and \hi profile linewidth \citep{2000ApJ...533L..99M, 2005ApJ...632..859M, 2016ApJ...816L..14L, 2019MNRAS.484.3267L}.  Consequently, and in response to our concern regarding the bias with faint galaxies in the TFR study  by \citet{2020ApJ...902..145K}, the same sample has been reanalyzed with the BTFR methodology by \citet{2022MNRAS.511.6160K}.

The BTFR requires the additional component of \hi fluxes, an observable acquired simultaneously with linewidths.  The need for a robust \hi detection gives focus to the condition that the sample is \hi flux limited: photometry for any target with sufficient \hi flux is easily obtained.  However, an \hi flux limit translates to a cut in gas mass that increases with distance.  This trend results in a bias that must be addressed in order to obtain distance measures of value. \citet{2022MNRAS.511.6160K} developed a procedure that was demonstrated with mock data to provide unbiased distance estimates.

Application of the BTFR also requires the translation of luminosities into approximations of stellar mass, involving color terms.  The conversions are linear, but uncertainties are further compounded by the summation of stellar and gas mass components.  On the one hand, there is a greater complexity with the BTFR, but on the other hand, the linkage with the dark matter-dominated total mass is expected to be tighter.  Scatter as evaluated from $H_i$ values at $V>4,000$~\kms\ is 22\% in distances, comparable to that with the TFR. \citet{2022MNRAS.511.6160K} provide BTFR distance estimates for 9967 galaxies.


\begin{figure}[]
\centering
\includegraphics[width=0.48\textwidth]{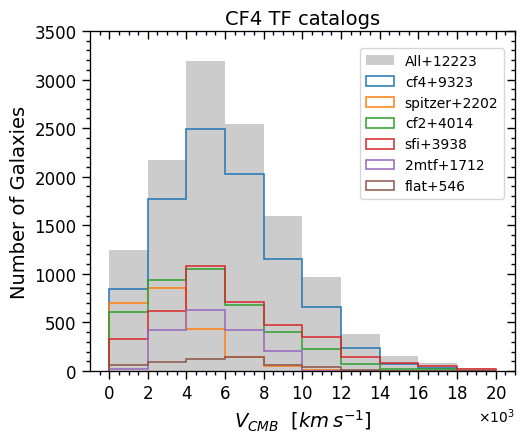}
\caption{
Cumulative histogram of TF targets with systemic velocity and a breakdown by subsample as given by the legend. \label{fig:histv_tf}
}
\end{figure}

\begin{figure*}[!]
\centering
\includegraphics[width=0.99\textwidth]{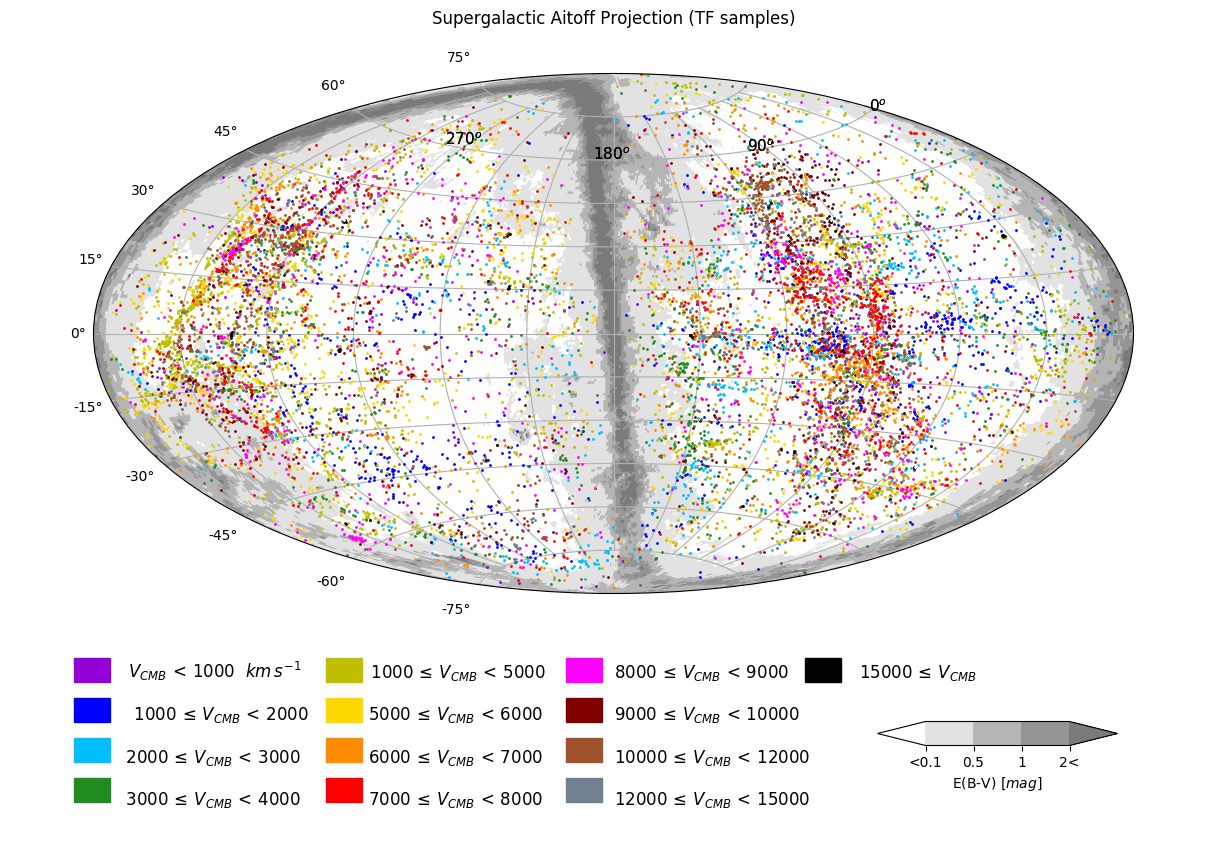}
\caption{
An Aitoff projection in supergalactic coordinates of the distribution of the 12,223 galaxies constituting an ensemble of TFR samples.  Colors relate to systemic velocities of the group of a galaxy as given in the table below the map.  Milky Way extinction levels are cast in shades of gray.  The dense roughly vertical swaths of objects in both supergalactic hemispheres lie in the decl. band of the Arecibo Telescope.  
\label{fig:aitoff_tf}
}
\end{figure*}

\subsection{Ensemble of TFR and BTFR Sources}

{\it Cosmicflows-4} assembles TFR distances to $12,412$ galaxies, the largest, most coherent compilation to date by this methodology.  The most important contribution (9,967 galaxies) is the new BTFR sample discussed by \citet{2022MNRAS.511.6160K} (hereafter cf4).  This new sample is compared and merged with five TFR samples: the assembly of 5,980 cases in {\it Cosmicflows-2} that itself is broken into a part (4,069) derived within our collaboration (hereafter cf2) and a part (3,957) emanating from the SFI++ study \citep{2007ApJS..172..599S} (hereafter sfi), 2,251 galaxies discussed in {\it Cosmicflows-3} incorporating photometry from Spitzer Space Telescope images (spitzer), 1,715 galaxies utilizing 2MASS photometry (2mtf) \citep{2019MNRAS.487.2061H}, and 551 extreme edge on galaxies \citep{2018MNRAS.479.3373M} (flat). 

Figure~\ref{fig:histv_tf} is a histogram of the run of velocities for the TF subsamples and the full TF sample.
The distribution of the combined TF sample is displayed in supergalactic coordinates in Figure~\ref{fig:aitoff_tf}.  The bands of high object density crossing the two supergalactic hemispheres lie in the $0<\delta<38$ decl. zone accessed by the Arecibo Telescope.      

Our analysis began with each sample alone.  At systemic velocities above $~4,000$~\kms\ cosmic expansion velocities are expected to overwhelmingly dominate deviant velocities.  Hence, a necessary (not sufficient) criterion a sample should satisfy is approximate constancy in the Hubble parameter for individual galaxies, $H_i=f_icz_i/d_i$, averaged in velocity bins.  The results of this test for all but the most recent flat galaxy sample were presented in \citet{2020ApJ...902..145K}.  A significant drift toward smaller $\langle H_i \rangle$ (larger derived distances) was evident in the 2mtf sample.  An adjustment to negate this trend was introduced by \citet{2020ApJ...902..145K}, see \S 5, and is incorporated in the current work.  The flat sample passes the $\langle H_i \rangle$ constancy test.  Note that  absolute $\langle H_i \rangle$ values are not an issue at this stage; they can be (and are) different for each sample. 

This test of the constancy of $H_i$ with redshift provides as a side product an evaluation of the rms dispersion in measurements within each sample. The measured values include dispersion in velocities and intrinsic dispersion but these components are unimportant if, as we do, we restrict attention to velocities greater than 4,000~\kms.  We find the following characteristic rms dispersions for seven samples (treating separately the optical and infrared components of cf4 (whence, cf4-op and cf4-ir), the two components of {\it Cosmicflows-2}, cf2 and sfi, and the components of spitzer with and without color corrections (spitzer-cc and spitzer-nc).  The rms dispersions for the BTFR cf4-ir and cf4-op are 0.45 and 0.47 mag respectively, while for the two-parameter TF studies dispersions are 0.40 for all three cf2, sfi, and spitzer-cc, 0.50 for both spitzer-nc and 2mtf, and 0.55 for flat.  Distance values for 2mtf at Local Group frame velocities less than 2000~\kms, as evaluated by the Hubble parameter test and the test to be discussed next, are systematically too low and we reject all those 2mtf measurements.

Another test of the samples applied by \citet{2020ApJ...902..145K} was to focus on differences in distance moduli between cf4 and an alternate sample: $\langle \mu_{cf4} - \mu_{alt} \rangle$ where $alt$ is any of the other samples (now extended to include flat).  This test is particularly useful for the isolation of egregiously bad distance values in one of the samples.  Much less than 1\% of cases in cf2, sfi, spitzer, and cf4 are rejected by this test. With 2mtf $\sim 2\%$, and with flat $\sim 5\%$, are rejected.

We now turn our attention to the integration of these samples into a global maximally consistent compilation of TFR distances.


\subsection{Preliminaries}
\label{sec:prelim}

Our goal is to combine the distinct TFR subsamples into a single TFR sample.  Before this integration, each subsample has its own zero-point scaling.  Here, we revise the zero-points of subsamples to achieve statistical equality between them. We stress that we do not make relative changes in moduli {\it within} a subsample in this process. Doing so would subvert the utility provided by multiple subsamples in reducing systematics. 

It is evident that TFR samples have non-Gaussian outliers. Steps have been described to remove strongly deviant cases but our initial integration of subsamples reveals additional instances.  Applying a $3.5\sigma$ rejection criterion caught 275 cases among 22,233 measures (1.2\%) where there would be nine with a normal distribution. These outliers are removed.


Next, we want to profit from the advantages of averaging over groups discussed in \S \ref{sec:groups}.  We begin by weighted averaging of the distance moduli of all galaxies within a 1PGC group within a single subsample.  Individual weights are formed from the inverse square of rms uncertainties. This average and associated weight is one object in the ensuing analysis. 
Accordingly, each subsample is reduced to a quantity of objects (halos, groups, clusters) composed of from one to many individual galaxies, each identified by a 1PGC number.



\begin{figure*}[t]
\centering
\includegraphics[scale=0.45]{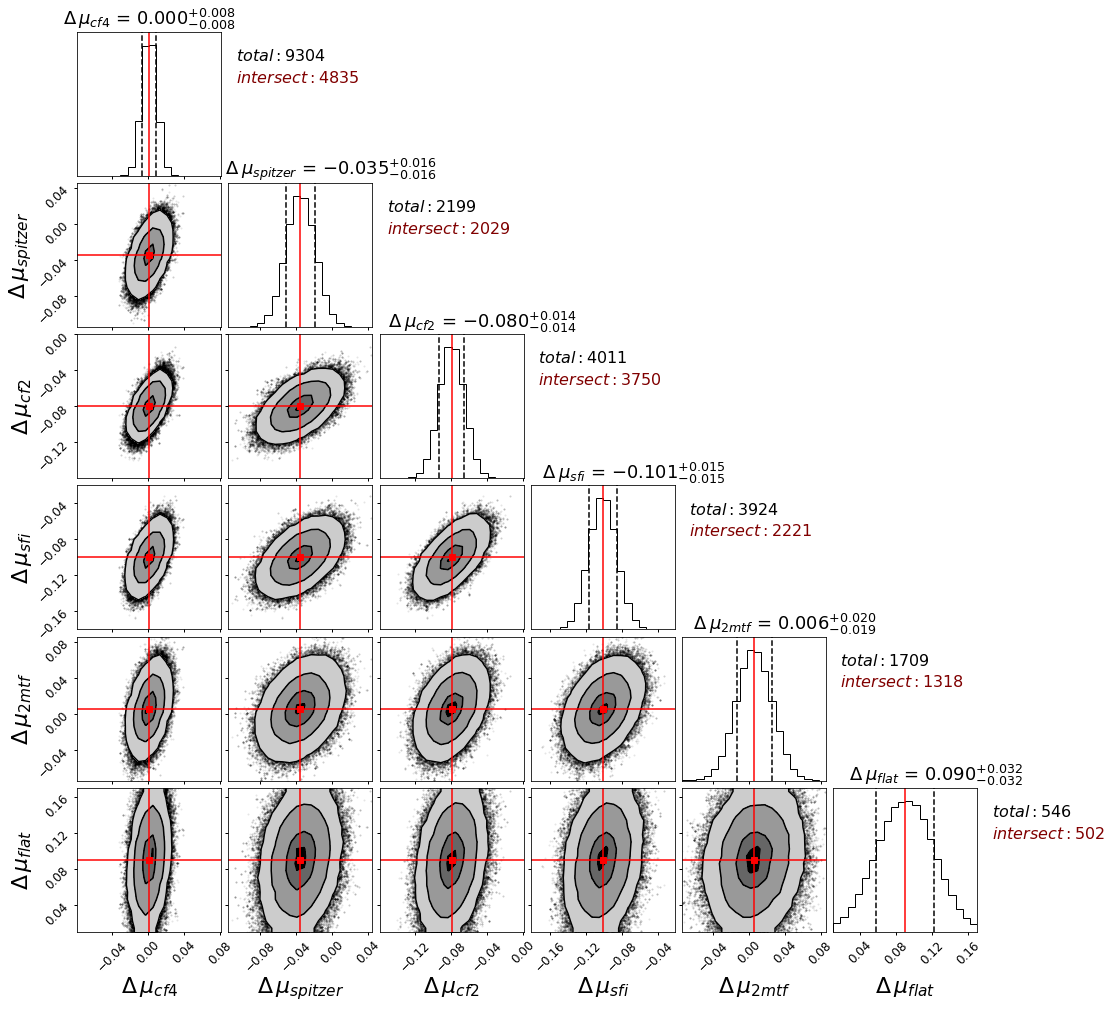}
\caption{
The posterior distribution of the optimized zero-points of TFR catalogs with respect to cf4. Contours represent $\sigma/2$, $\sigma$, $3\sigma/2$ and $2\sigma$ levels of the two-dimensional distributions and they enclose 12\%, 39\%, 68\%, and 86\% of the distributed points, respectively. 
Two vertical dashed lines in each of the one-dimensional histograms specify the region that accommodates 68\% of the points, and the red vertical line identifies the median of the distribution.
Each panel covers $\pm0.08$ mag about the center of the distribution.
\label{fig:mcmc_tf}
}
\end{figure*}


\subsection{Combining all TFR distances: Bayesian approach}
\label{sec:bayesianMerge}

Ultimately, we want to merge all samples by all methodologies into a coherent set with a zero-point established by geometric distance measurements.  At this stage, it is sufficient to bring all TFR subsamples onto a common scale. The baseline TFR scale will be set by our new cf4 subsample that should lie close to our final scale, given its linkage to Cepheid and TRGB measures as discussed by \citet{2020ApJ...896....3K, 2020ApJ...902..145K, 2022MNRAS.511.6160K}.

Here, we pursue our goal to find the global modulus offset of each sample, ``s", from that of cf4, where ``s" stands for any of the samples we introduced earlier in this chapter (cf2, sfi, spitzer, 2mtf, flat).
We will adjust the reported distance moduli within each sample, $DM_{in}^{(s)}$, following $DM^{(s)} = DM_{in}^{(s)} + \Delta \mu_{s}$ in order to set all cataloged distances on the same scale. By our convention, $\Delta \mu_{cf4}=0$.
We treat these adjusting values as a set of free parameters that are optimized together in a Bayesian framework. The best offset parameters minimize the total deviation of adjusted object distance moduli (groups and individual galaxies) from the weighted distance modulus averages offered by all samples together. 

Our objective is to find the posterior probability distribution $\Pc(\Theta|\mathcal{D})$, with $\Theta$ being the vector of all moduli offsets, ($\Delta \mu_{s1}$, $\Delta \mu_{s2}$, ...). $\Dc$ holds the original cataloged distance moduli, $DM_{in}^{(s)}$. According to conditional probability theory, $\Pc(\Theta|\mathcal{D}) \propto \Pc(\mathcal{D}|\Theta)\Pc(\Theta)$. Having no prior knowledge about the distribution of the moduli offsets implies $\Pc(\Theta)=1$ and subsequently $\Pc(\Theta|\mathcal{D}) \propto \Pc(\mathcal{D}|\Theta)$, where the right-hand side is the likelihood function, $\Lc$.
We assume that all measured object distances are independent with Gaussian uncertainties. Therefore, for each object, $n$, the likelihood function is the multiplication of a set of independent probabilities given as  
\begin{equation}
\label{Eq:likelihood}
\Lc_n= \prod_{All ``s"} \frac{1}{\sqrt{2 \pi \sigma_{n,s}^2}} ~\exp\frac{-1}{2}\Big(\frac{DM_n^{(s)} - \langle DM \rangle_n}{\sigma_{n,s}}  \Big)^2 ~,
\end{equation}
iterating over all distance catalogs. 
$\langle DM \rangle_n$ is the weighted average distance modulus of the $n^{th}$ object that is derived from the adjusted distance moduli, $DM_n^{(s)}$ is the distance modulus of the object in the sample ``s", and $\sigma^2_{n,s}$ is the variance of $DM_n^{(s)} - \langle DM \rangle_n$, which is determined by adding the uncertainties of the associated parameters in quadrature. Likewise, the total likelihood function for all objects is $\Lc_{tot}=\prod_{n=1}^N\Lc_n$, where $N$ is the total number of objects (groups and individuals).
It is simpler to work with the logarithm of the likelihood function, which is expressed as 

${\rm log} \Lc_{tot} = -\sum_{n=1}^N \chi_n^2/2 $, where 
\begin{equation}
\chi^2_n = \Big(\frac{DM_n^{(s)} - \langle DM \rangle_n}{\sigma_{n,s}}  \Big)^2~.
\end{equation}
Adopting a flat prior distribution for the moduli offsets leaves us with a $\chi^2$ minimization problem. We are interested in a set of moduli offsets that minimizes $\chi^2_{tot}=\sum_{n=1}^N \chi_n^2$.

To sample the posterior distribution, $\Pc(\Theta|\mathcal{D})$, we use the Python package {\it emcee} \citep{2013PASP..125..306F}, which implements Markov chain Monte Carlo (MCMC) simulations to explore the parameter space. Starting from our likelihood function, we generate 128 chains each with the length of 10,000. We remove the first 1,000 steps which are conservatively chosen to ensure that the remaining steps adhere to Markov chain statistics. Figure~\ref{fig:mcmc_tf} illustrates the corner plots for the resulting posterior distribution of $\Delta \mu_{s}$. The topmost panel of each column shows the one-dimensional distribution of the corresponding sampled parameter, overlaid with the median values (red solid line) and the lower/upper bounds corresponding to 16/84 percentiles (black dashed line). Horizontal and vertical red lines in the two-dimensional distributions exhibit the location of the median values that are adopted as the optimum moduli offsets of the corresponding catalogs with respect to cf4.

The variance for a given subsample that is recorded in Fig.~\ref{fig:mcmc_tf} depends on both the uncertainties in individual measurements and the number of intersections with other subsamples.  The individual uncertainties between the alternate TF subsamples are only modestly different, so it is the numbers of intersections that dominate. 


\section{FP Distances to Early-type Galaxies}
\label{sec:FP}

The FP methodology \citep{1987ApJ...313...42D, 1987ApJ...313...59D}, with its applicability to early-type galaxies, provides a complement to the TFR.  The accuracies of individual measurements are comparable. While the gas-rich systems observed with the TFR are widely dispersed, the old star-dominated systems favored for FP observations tend to clump in regions of high density.

Here, in {\it Cosmicflows-4} we combine results from five programs.  Three of these were already included in {\it Cosmicflows-2}: contributions for a total of 1508 galaxies to be referred to as smac \citep{2001MNRAS.327..265H}, efar \citep{2001MNRAS.321..277C}, and enear \citep{2002AJ....123.2990B}.  Individually these sources provide distances for 690, 696, and 447 galaxies, respectively.
Contributions from a fourth program, 6dFGSv \citep{2014MNRAS.445.2677S} were included in {\it Cosmicflows-3}.  This sample of 7,099 galaxy distances is particularly important as the numerically dominant source of distances in the celestial south.  However, by far the largest sample containing 34,059 galaxies is a new contribution restricted to the celestial and galactic north that draws on data extracted from the SDSS.
The three earliest FP surveys, smac, efar, and enear provide valuable bridges across the celestial hemispheres, and are important given there is only a slight overlap (41 cases) between the 6dFGSv and SDSS samples. 


\subsection{The 6dFGSv Sample}
\label{sec:6dF}

While the 6dFGSv sample was originally included in {\it Cosmicflows-3}, in this latest work we provide a new recalibration of this sample based on the findings of \cite{Qin2018} designed to explore and remove spurious flows. In total, the 6dFGSv sample subtends the entire $\delta<0^{\circ}$ sky, except for regions with galactic latitude $|b|<10^{\circ}$.  Its 8,885 objects incorporate many of the brightest early-type galaxies in 6dFGS \citealt{2009MNRAS.399..683J}), nominally selected to have a spectral signal-to-noise ratio $>5$, total $J$-band magnitude $< 13.65$, redshift $cz<16,500$~\kms\ and velocity dispersion greater than $112$~\kms\ \citep{2014MNRAS.443.1231C}. Further refinements to this selection include visual classification and removal of galaxies based on their morphological type (although as demonstrated in \citet{2016AJ....152...50T} not all remaining galaxies are classified as ellipticals) and removal of objects with undesirable spectral features or poor spectral template fits \citep{2014MNRAS.443.1231C}. Photometry for the Fundamental Plane sample was obtained by cross-matching with the 2MASS survey \citep{2000AJ....119.2498J, 2006AJ....131.1163S}.

\cite{2014MNRAS.445.2677S} produce peculiar velocity measurements with this sample by modeling the logarithmic difference in observed and cosmological distances to each galaxy as a function of the logarithmic difference between their observed effective radii and the effective radii predicted from the best-fit FP. For each galaxy, one can compute the probability of it having a particular distance modulus by assuming a Gaussian probability distribution function (PDF) about the FP. However, this procedure is complicated by Malmquist bias and the selection function of the 6dFGSv data, particularly the magnitude limit. The presence of a magnitude limit cuts a slice through the FP, such that the PDF is no longer normalized. Because the magnitude limit is in apparent magnitudes, the portion of the Fundamental Plane that cannot be observed varies with distance, and so the normalization of the PDF for each galaxy also depends on distance. To counteract this effect, \cite{2014MNRAS.445.2677S} produced a calculation of this normalization as a function of distance using simulations drawn from the best-fit 6dFGSv FP with a $J<13.65$ limit.

In subsequent work by \cite{Qin2018}, similar simulations reproducing the FP, selection function, and methodology applied to the 6dFGSv data were used to measure the bulk flow. A significant offset between the measured and true bulk flows in the simulations was identified in the direction directly toward the southern celestial pole. It was found to be possible to remove this effect in the simulations (and subsequently the data) by recalculating the normalization of the probability distribution for each galaxy using an ad hoc, brighter, magnitude limit of $J<13.217$. In this work we repeat this calculation, paying special attention to not only the bulk flow, but also the Hubble parameter in radial shells.

We start by generating 128 mock 6dFGSv surveys matching the methods in \cite{2012MNRAS.427..245M} and \citet{Qin2019}, which are then run through a reconstruction of the 6dFGSv pipeline using a magnitude limit of $J<13.65$. Unlike previous works, the normalization of the PDF for each galaxy is computed using numerical Monte Carlo integration of the truncated 3D Gaussian PDF, rather than summing over simulations. This procedure was found to result in less noise and was far more reliable for computing an accurate value at large comoving distances where by design the number of galaxies available to compute the probability, even over 128 simulations, quickly falls to zero. 

\begin{figure}[t]
\centering
\includegraphics[width=0.48\textwidth]{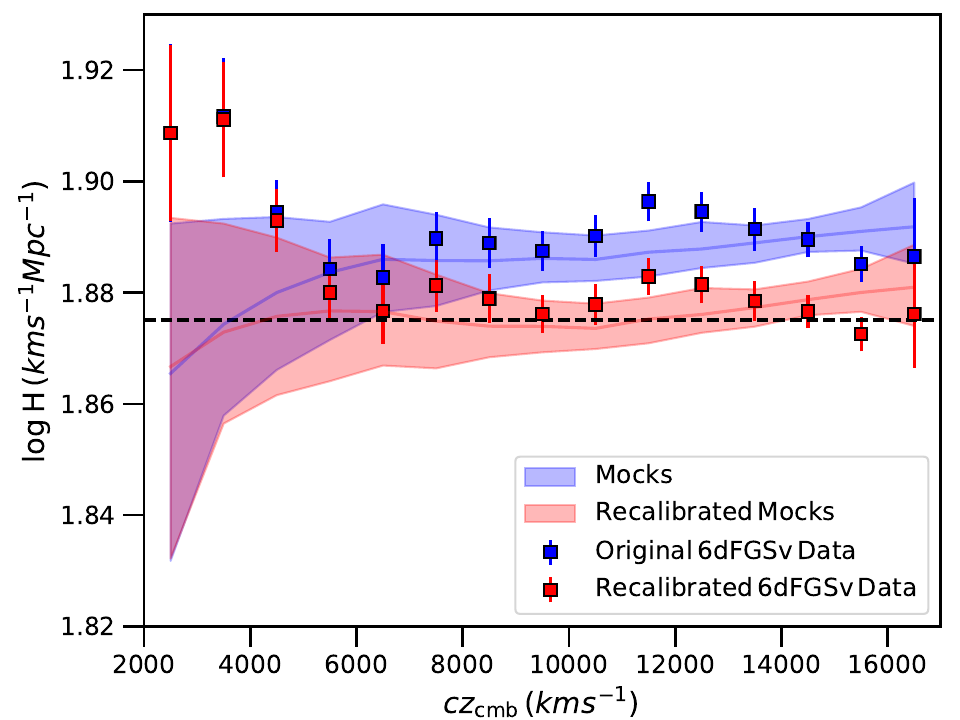}
\caption{Measurements of the weighted mean Hubble parameter in redshift bins of width 1000~\kms\ from the 6dFGSv data (points) and simulations (band). The bands show the median value and $68\%$ percentile region for the 128 mock realizations while the horizontal dashed line denotes the input value $H_{0}=75$~\kmsMpc\ used to compute the distance. We expect the Hubble parameter to be roughly constant with redshift and lie close to the input value. The blue band/points show the Hubble parameter using the original 6dFGSv methodology with a magnitude limit of $J<13.65$ for the Malmquist bias correction. The red band/points show the recalibrated results using $J<13.38$, which clearly reduces the bias seen in the mocks. Though high, the 6dFGSv data for $cz_{\mathrm{cmb}}<5000$~\kms\ are still within the $95\%$ region computed from the simulations.
\label{fig:6dFGSv_hubble}
}
\end{figure}

\begin{figure*}[t]
\centering
\includegraphics[width=0.48\textwidth]{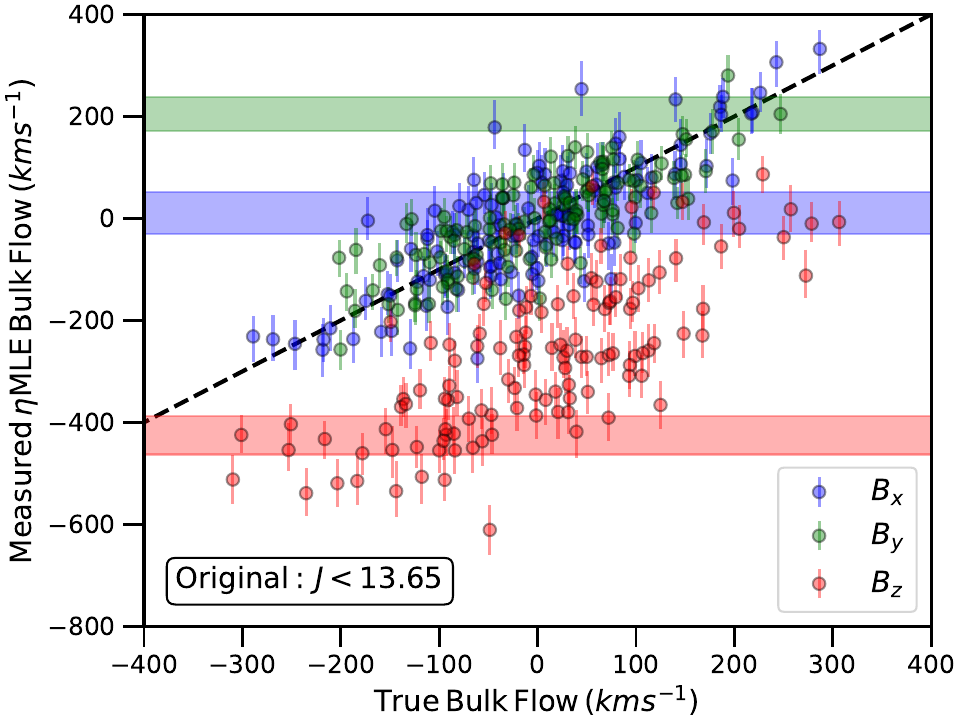}
\includegraphics[width=0.48\textwidth]{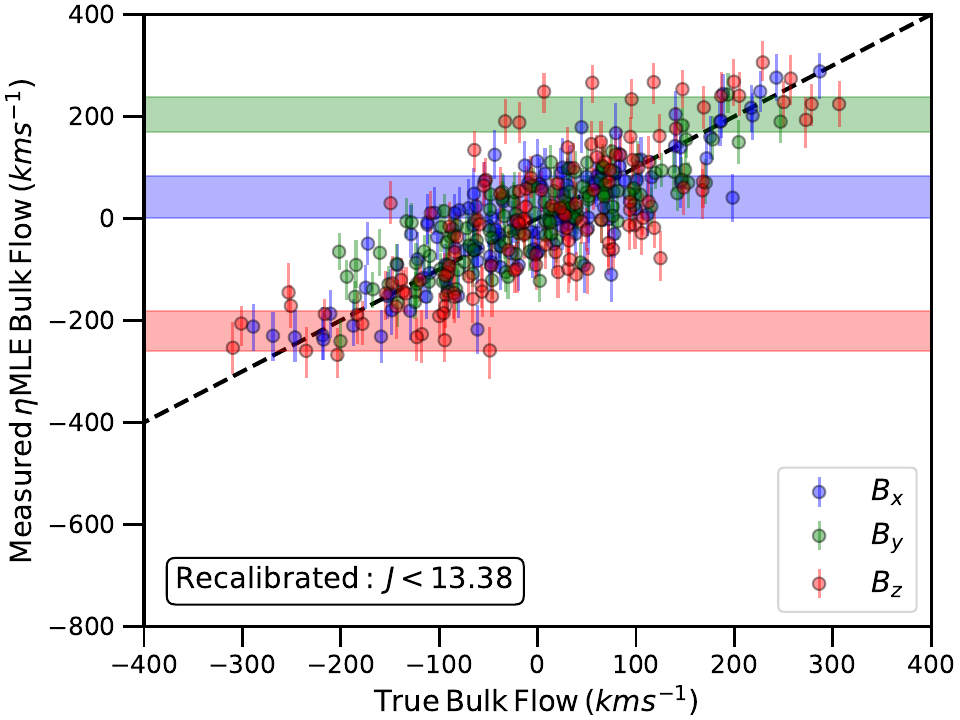}
\caption{Measurements of the bulk flow in each direction from 6dFGSv mocks (points) and data (bands) using the $\eta$-MLE method of \cite{Qin2018}. For each simulation, we plot the measured maximum likelihood bulk flow against the true bulk flow calculated by averaging over the true peculiar velocity of each galaxy. The true bulk flow for the data is not known \textit{a priori} and so the measurement is included as a horizontal band. In both cases, error bars/regions denote the equal likelihood bounds encapsulating $68\%$ of the posterior. The left-hand panel shows the results using the original 6dFGSv methodology and $J<13.65$ magnitude limit for the Malmquist bias correction/PDF normalization. The right-hand panel shows the recalibrated results using $J<13.38$. The recalibration removes the strong negative bias in the bulk flow in the direction of the south celestial pole ($z$-axis in this coordinate system) seen in the simulations and believed to also be present in the data.
\label{fig:6dFGSv_BF}
}
\end{figure*}

The bulk flow in each simulation was then computed using the $\eta$-maximum likelihood estimator \citep{Kaiser1988,Qin2018}, as was the weighted mean value of 
\begin{equation}
    \mathrm{log_{10}}(H_{i}) = \mathrm{log_{10}}\biggl(\frac{f_icz_i}{d_{i}}\biggl).
    \label{eq:logh}
\end{equation}
in redshift bins, where $f_i = 1+1/2[1-q_{0}]z_{i}-1/6[1-q_{0}-3q^{2}_{0}+j_{0}]z_{i}^{2}$, $z_{i}$ is the redshift of the galaxy, $q_{0}$ and $j_{0}$ are the acceleration and jerk parameters, and $c$ is the speed of light. Here, $d_{i}$ is the luminosity distance to each galaxy, computed assuming $H_{0} = 75$~\kmsMpc. For comparison, the same quantities are computed for each simulation using the true luminosity distance and peculiar velocity of each simulated galaxy.

The results of this procedure are shown in Figures~\ref{fig:6dFGSv_hubble} and ~\ref{fig:6dFGSv_BF}. Also plotted alongside are the results for the original 6dFGSv data. From the binned Hubble parameters, it is clear that the mocks with distance moduli computed using the 6dFGSv pipeline exhibit an outflow and do not lie on the expected $H_{0} = 75 ~\kmsMpc$ line. The distribution of these same mocks matches the data, which leads us to conclude the same is likely true for the data too. This trend is not obvious without the presence of simulations (and so not highlighted previously), particularly because cosmic variance at $cz_{\mathrm{cmb}}<5000$~\kms\ seems to be scattering the observed Hubble parameters high in the data. 

The effect on the bulk flow is particularly pronounced. It was found by \cite{Qin2018} that the measured bulk flow in the simulations is biased quite negatively in the direction of the southern celestial pole compared to the bulk flow known to exist in the simulations. Although the {\it true} bulk flow in the 6dFGSv is not known and can only be estimated, the measured value is consistent with the biased mock results, leading us to conclude that the data is similarly biased.

As with the previous work by \cite{Qin2018}, we correct for this problem by modifying the magnitude limit used in the Malmquist bias correction/normalization of each galaxy's PDF. By iterating, a magnitude limit of $J<13.38$ was found to produce binned Hubble parameters that are flat with redshift while also substantially reducing the difference between the adjusted and measured bulk flows. The results of applying this limit to the mocks and data are shown in Figs.~\ref{fig:6dFGSv_hubble} and ~\ref{fig:6dFGSv_BF}. The small differences between the optimal value found here and that used in \cite{Qin2018} are likely the result of the more rigorous calculation of the normalization using numerical integration adopted in this work.

We believe this recalibration to be robust and so use the updated 6dFGSv data in {\it Cosmicflows-4}. The source of the discrepancy between the magnitude limit used to construct the simulations and the optimal value found for the Malmquist bias correction is unclear, but indicates a discrepancy between the best-fit 6dFGSv FP (from which the simulation apparent magnitudes are derived) and the assumed magnitude limit of the data. It is not inconceivable that the true magnitude limit of the 6dFGSv data is in reality brighter than the nominal selection function, particularly in light of the other aspects of the sample selection required to go from the 2MASS photometry and 6dFGS spectra to the FP sample, and then again to the peculiar velocity sample. However, there may be more to the picture. A preferable solution would be to perform a joint fit for the FP parameters and peculiar velocities simultaneously, again validated against simulations; however, such an analysis is beyond the scope of the current work.

\begin{figure}[!]
\centering
\includegraphics[width=0.48\textwidth]{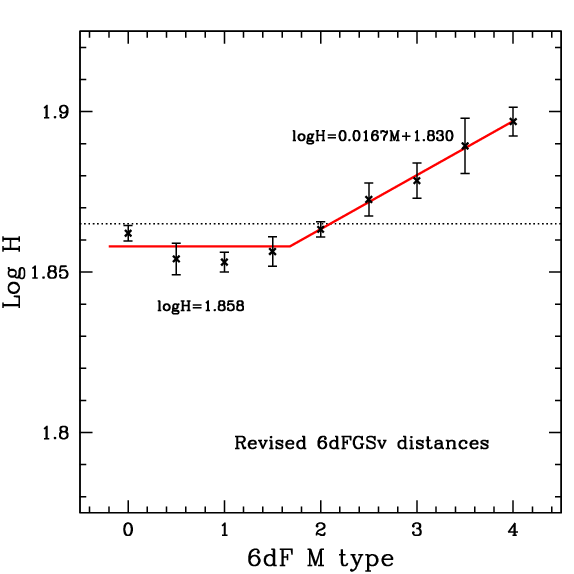}
\caption{Trend in values of log$H_i = {\rm log}(f_iV_i/d_i)$ with morphological $M$ type.  Galaxies with M$>2$ increasingly manifest the properties of disk systems.
}
\label{fig:Mtype}

\end{figure}

The systematic problem as a function of morphological type identified by \citet{2016AJ....152...50T} remains in the revised bias-adjusted 6dFGSv distances.  Candidates in the 6dFGSv compilation are given a morphology $M$ description, with $M=0$ for ellipticals, $M=2$ for lenticulars, and $M=4$ for spirals. As seen in Figure~\ref{fig:Mtype}, there is a clear drift in $\langle f_iV_i/d_i \rangle$ with $M$, where $V_i$ and $d_i$ are individual galaxy velocities and distances and $f_i$ is defined in association with Eq.~\ref{eq:logh}.  The drift is in the sense that $d_i$ values are increasingly measured too low with increasing $M$.

This situation is not too surprising given that the FP pertains to galaxy bulges.  The admixture of disk contributions evidently creates a systematic.  The revisions discussed above to 6dFGSv distances has not addressed this morphology-related problem. Adjustments are made to distances that statistically counter the trend shown in Fig.~\ref{fig:Mtype}.  Only the fitting parameters are changed from the adjustments made in \citet{2016AJ....152...50T}.  The distance moduli accepted into {\it Cosmicflows-4} incorporate both the revised bias corrections and the morphological corrections described in this section. 


\subsection{FP with SDSS Photometry and Spectroscopy}
\label{sec:sdss}

A new addition to the \textit{Cosmicflows} program is the SDSS peculiar velocity catalog \citep{2022MNRAS.515..953H}. Containing a total of 34,059 objects, it is currently the largest single source of extragalactic distance measurements available. It also contributes, along with a modest number of supernova observations, the most distant objects in the \textit{Cosmicflows-4} catalog, extending up to a maximum cosmic microwave background (CMB)-frame redshift of $z_{\mathrm{CMB}}=0.1$. These measurements were made by combining public photometric data from SDSS Data Release 14 \citep{2018ApJS..235...42A} over a contiguous $\sim7,000\,\mathrm{deg}^{2}$ area with existing spectroscopic H$\alpha$ and velocity dispersion measurements \citep{2013MNRAS.431.1383T}.

The full catalog is magnitude limited to $10.0 \le m_{r} \le 17.0$ and velocity dispersion limited to $\sigma > 70\,\mathrm{km\,s^{-1}}$. The selection of the sample also included a number of additional cuts on the surface brightness profiles, concentration index, color, axial ratio, H$\alpha$ equivalent width, and visual morphology to ensure a clean sample of elliptical galaxies is retained. As part of this process, cuts were applied using the morphological $M$-type \citep{2011A&A...529A..53T} \textit{before} fitting of the FP, so no correction of the form seen in Fig.~\ref{fig:Mtype} for the 6dFGSv sample is required for the new SDSS data.

After data cuts, the sample was first fit using the 3D Gaussian FP model. Then, for each galaxy, the PDF of the logarithmic distance ratio between cosmological and observed comoving distances was obtained. 
The normalization of this PDF is set by integrating over the portion of the FP within which each galaxy would be observable at a given cosmological distance, and hence encodes the selection bias. Larger proposed distances for each galaxy end up being up-weighted to account for the increased cosmological volume in which a galaxy could be found, and the lower sample completeness at those distances. Previous work (i.e., \citep{2014MNRAS.445.2677S}) computed this normalization using Monte Carlo simulations, which requires assuming the same normalization for each galaxy. However, \citet{2022MNRAS.515..953H} demonstrated it can also be computed numerically, which makes the calculation much faster and allows one to use a different normalization for each object.
The mean, standard deviation, and skew of each PDF were used to describe the full PDF of each galaxy.

Using cosmological simulations of the SDSS data and selection function, \citet{2022MNRAS.515..953H} demonstrated the measurements were unbiased as a function of redshift and absolute magnitude. However, they did identify a bias between the distance and group richness. This bias arises from a correlation between the mean surface brightness of galaxies in the sample and their corresponding group richness, possibly hinting at intrinsic correlations in the FP due to a fourth unmodelled parameter such as stellar age. This group properties bias was corrected in the catalog by fitting separate FPs to subsamples of the data as a function of group richness. These {\it corrected} distance measurements are the ones used in this work.

Calibration of the zero-point in the SDSS PV catalog is made difficult due to the small sky area. In \citet{2022MNRAS.515..953H} this calibration was performed by cross-matching to groups containing galaxies with \textit{Cosmicflows-3} distances. The use of groups was found to be important considering the group richness bias discussed above, although consistency with \textit{individual} measurements in the two catalogs was also demonstrated as long as the {\it uncorrected} SDSS distances were used.



\begin{figure*}[!]
\centering
\includegraphics[width=0.80\linewidth]{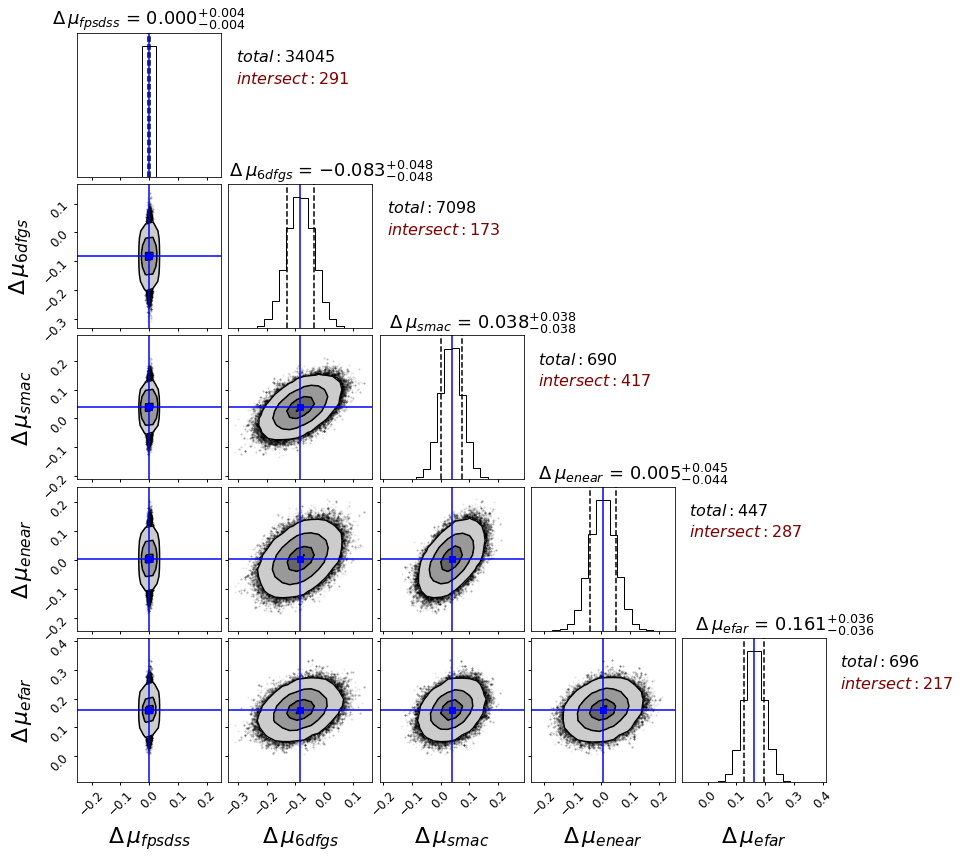}
\caption{
The posterior distribution of the optimized zero-points of FP catalogs with respect to SDSS. 
Each panel covers $\pm0.25$ mag about the center of the distribution. Other details are as in Figure \ref{fig:mcmc_tf}. 
\label{fig:mcmc_fp}
}
\end{figure*}

\begin{figure}[]
\centering
\includegraphics[width=0.48\textwidth]{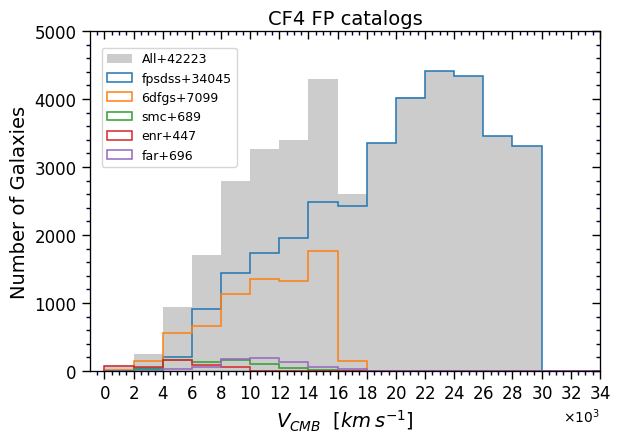}
\caption{
Cumulative histogram of FP targets with systemic velocity and a breakdown by subsample as given by the legend. 
\label{fig:histv_fp}
}
\end{figure}

\begin{figure*}[!]
\centering
\includegraphics[width=0.83\textwidth]{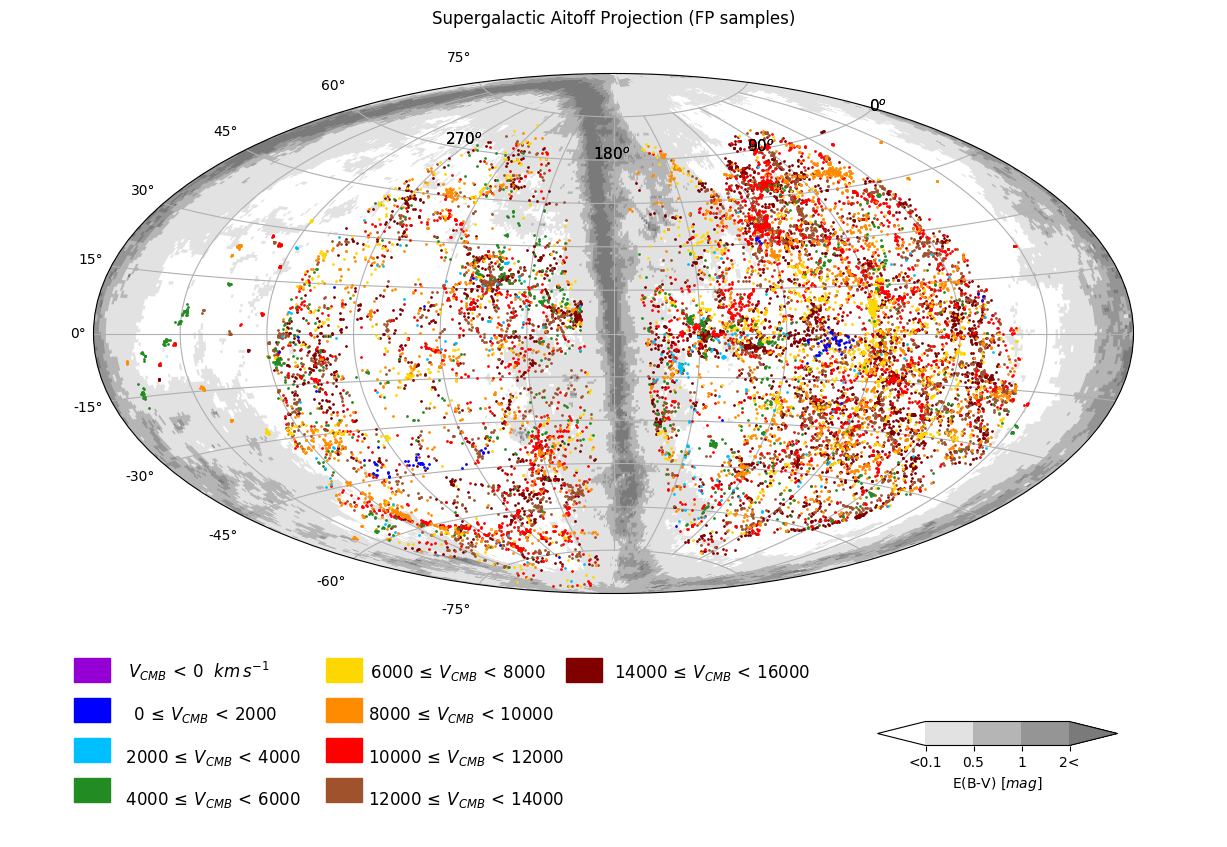}
\includegraphics[width=0.83\textwidth]{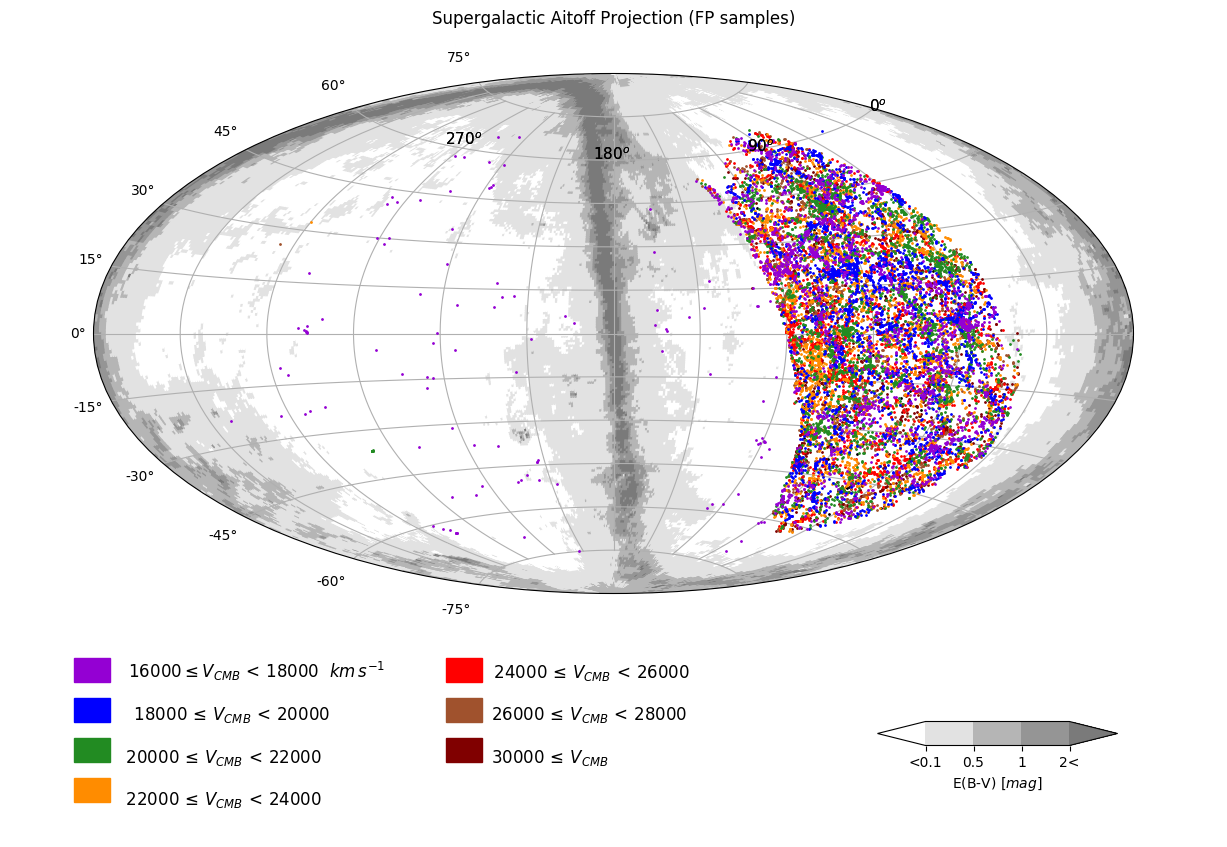}
\caption{
Aitoff projections in supergalactic coordinates of the distribution of the ensemble of FP samples.  Top panel: 16744 cases with $V<16,000$~\kms. Bottom panel: 25479 cases with $V>16,000$~\kms.  Colors relate to systemic velocities of the group of a galaxy as given in the table below the map.  Milky Way extinction levels are cast in shades of gray.  The dense roughly vertical swaths of objects in the north supergalactic hemisphere in both panels lie in the SDSS decl. band.  Contributions from the 6dFGSv sample lie to the left of the SDSS sample in the top panel.
\label{fig:aitoff_fp}}
\end{figure*}

\subsection{Combining all FP distances}
\label{sec:allFP}

Just as multiple acquisitions of a distance to a spiral galaxy with photometry and rotation curve information are highly correlated, so it is the case with multiple applications of the Fundamental Plane to early-type galaxies.
Hence, as an initial step we bind all the FP subsamples into a joint sample.
The same methodology as discussed in \S \ref{sec:bayesianMerge} is adopted to combine the distances of the FP catalogs of this study: SDSS, 6dFGSv, smac, enear and efar. 

As noted in the introduction of \S\ref{sec:FP}, there is only the tiniest overlap of 41 cases between 6dFGSv, which is entirely restricted to negative declinations, and SDSS, which is limited to the north galactic and celestial pole caps, dipping slightly into the celestial south to $\delta = -3.7^{\circ}$. Consequently, the smac, enear and efar samples play the important roles of providing scaling links between the two large samples.  Specifically, smac provides 118 and 90 links with SDSS and 6dFGSv respectively, enear provides 75 and 41 respectively, and efar provides 28 and 41 respectively.  Collectively, these three smaller samples provide 367 overlaps with SDSS and 200 overlaps with 6dFGSv.  

The 3.5$\sigma$ rejection criterion is used to identify two outliers in enear, one in smac and 17 in the 6dFGSv subsamples. These cases are removed prior to the subsequent calculations.

With the interlacing of FP samples, the SDSS distances are taken as the reference of comparisons, whence $\Delta \mu_{SDSS} = 0$ by our convention. 
The posterior distribution of the offset parameters ($\Delta \mu_s$) are explored with our MCMC procedure in a similar fashion to that explained in \S \ref{sec:bayesianMerge}. Figure \ref{fig:mcmc_fp} illustrates the posterior distributions of the moduli offsets of 6dFGSv, smac, enear and efar subsamples from SDSS.  It is to be appreciated that while the zero-point of the SDSS sample was set to be in approximate agreement with {\it Cosmicflows-4} distances \citep{2022MNRAS.515..953H}, it remains to receive further adjustment in the coupling with the full complement of methodologies described further along.  In particular, the union between methodologies afforded by group memberships will be exploited.

A histogram of the distribution of all FP measurements with velocity is shown in Figure~\ref{fig:histv_fp}.  The 6dFGSv and SDSS contributions are comparable at $V_{cmb} < 16,000$~\kms.  Only SDSS information is available between 16,000~\kms\ and the sample cutoff at $z=0.1$.  Figure~\ref{fig:aitoff_fp} shows the distribution of the galaxies with FP measurements on the sky, split into two panels for those above and below 16,000~\kms.


\section{SBF Distances to Early-type Galaxies}
\label{sec:sbf}

In predominantly ancient stellar systems, the brightest stars lie on the red giant branch where stellar envelopes inflate and hydrogen burning occurs in shells around inert helium cores \citep{2005essp.book.....S, 2017A&A...606A..33S}.  The very brightest of these stars, at the TRGB, are precursors to the onset of core helium burning and a restructuring of the star onto the HB.  The passband dependence of the luminosity of a star at the tip varies with metallicity, with enhanced molecular line blanketing pushing emission redward. At solar or super-solar metallicities, the stars at the TRGB peak in luminosity around 2~$\mu$m. The robustly characterized luminosity of the onset of core helium burning, and the fact that it is very bright, have been exploited to measure galaxy distances.  

In galaxies that can be resolved into individual stars, the preferred distance measurement  methodology uses properties of the TRGB, to be discussed in a later section.  For more distant targets, the stars blend, but the surface brightness appears mottled because of the Poisson statistics in the number of stars per resolution element, creating SBF \citep{1988AJ.....96..807T}.  The signal from the stellar fluctuations in the spatial Fourier power spectrum diminishes with distance as $1/d^2$, making it possible to measure galaxy distances with high accuracy with simple single-epoch imaging, which distinguishes SBF from other distance measurement techniques that rely on temporal monitoring (Cepheids and SNe) or spectroscopic observations (e.g., FP and TFR).

\subsection{Sources of SBF Measurements}

We have gathered SBF distances for 508 galaxies based on five sources of SBF measurements.  Two of these sources already found a presence in the earlier {\it Cosmicflows} compilations.  \citet{2001ApJ...546..681T} pioneered the methodology with observations from the ground at the optical $I$~band.  That program targeted 300 E/S0 galaxies including essentially all E galaxies within 2,000~\kms\ and a sampling out to 4,000~\kms.  The other earlier source of SBF distances was the HST study of E/S0 galaxies in the Virgo and Fornax clusters \citep{2007ApJ...655..144M, 2009ApJ...694..556B, 2010ApJ...724..657B}.  The exceptional spatial resolution of HST resulted in distance measurements with sufficient accuracy to resolve the three-dimensional structure of the Virgo Cluster, distinguishing the M, W, and W$^{\prime}$ background galaxy groups, and to provide an accurate differential distance between the Virgo and Fornax clusters. 

Now with {\it Cosmicflows-4} we add SBF distance information from two new sources.  The first of these \citep{2018ApJ...856..126C} is an extension of the HST Virgo Cluster study derived from the $u^{\star},g,i,z$ imaging Next Generation Virgo Cluster Survey carried out with the Canada-France-Hawaii Telescope \citep{2012ApJS..200....4F}.  Our knowledge of the Virgo Cluster environs, and particularly, of the separation of the W$^{\prime}$ group 50\% farther away in projection and the $W$ and $M$ structures twice as faraway, are given improved clarity with this enhanced distance compilation.

The second new SBF source, infrared SBF with the HST WFC3/IR camera, is a harbinger of a particularly promising improvement to the SBF methodology \citep{2021ApJS..255...21J}. The wide field of view and improved sensitivity of WFC3/IR over earlier ground-based and NICMOS IR observations makes it an optimal instrument for SBF measurements. Red giant stars are particularly bright in the near-IR, making IR SBF much brighter and easier to measure than at optical wavelengths, especially from space where the sky background is greatly reduced. \citet{2015ApJ...808...91J} established the calibration for SBF in the F110W and F160W filters using observations of 16 Virgo and Fornax galaxies. As discussed in \citet{2021ApJ...911...65B}, these IR SBF distances are  tied to the $I$-band SBF observations \citep{2009ApJ...694..556B}, the Cepheid distances to Virgo and Fornax, and the LMC distance modulus of 18.477 \citep{2019Natur.567..200P}.

\begin{figure*}[!]
\centering
\includegraphics[width=0.7\linewidth]{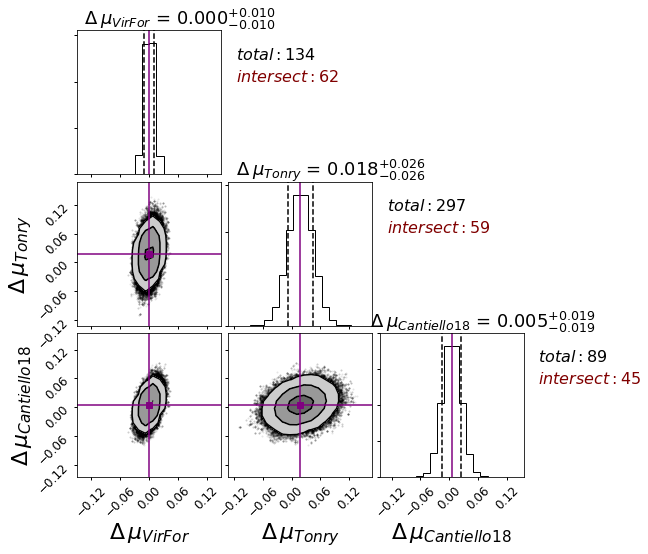}
\caption{
The posterior distribution of the optimized zero-points of SBF catalogs with respect to the Virgo-Fornax catalog. 
Each panel covers $\pm0.15$ mag about the center of the distribution. Other details are as in Figure \ref{fig:mcmc_tf}. 
\label{fig:mcmc_SBF}
}
\end{figure*}

The HST F110W SBF observations reach a distance of 80 Mpc in a single orbit, with an observational error of 4--5\% \citep{2021ApJ...911...65B,2021ApJS..255...21J}. A number of different WFC3/IR F110W programs have now been mined for usable observations. Two of these programs targeted galaxies specifically for SBF distance measurements. The program MASSIVE seeks to understand the stellar dynamical properties and central black hole masses of massive galaxies within $\sim100$~Mpc \citep{2014ApJ...795..158M}. 
This study includes the determination of 41 high-quality IR SBF distances. 
A second targeted SBF study measured distances to 19 early-type type Ia supernova (SN Ia) host galaxies 
\citep{2022arXiv220412060G} (P. Milne et al.\ 2022, in preparation). These data sets, along with independent observations of NGC~4874 in Coma \citep{2016ApJ...822...95C, 2017AAS...22914305B} and NGC~4993, the 2017 gravitational wave event host \citep{2018ApJ...854L..31C}, sum to a total of 62 galaxies out to 100~Mpc for which high-precision distances are now known.


\begin{figure}[]
\centering
\includegraphics[width=0.48\textwidth]{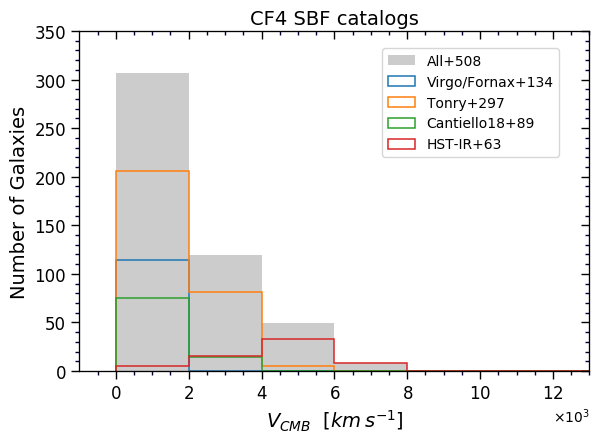}
\caption{Cumulative histogram of SBF targets with systemic velocity and a breakdown by subsample as given by the legend. \label{fig:histv_sbf}
}
\end{figure}



\subsection{Combining SBF subsamples}
\label{sec:allsbf}

There is significant overlap between only three of the four SBF subsamples.  The new IR SBF contribution has an overlap of only eight targets with Tonry.  
The corner plot for the three SBF subsamples that can be compared is seen in Figure~\ref{fig:mcmc_SBF}.  The Virgo/Fornax collection based on HST optical band observations is taken as the reference.  Whereas with the TF and FP studies galaxy-group comparisons help reduce errors, in the case of SBF the galaxies are relatively nearby and the individual accuracies are greater so comparisons are strictly galaxy-galaxy.  The integration of the full four SBF contributions awaits the global MCMC compilation of all methodologies.

Figure~\ref{fig:histv_sbf} is a histogram of the cumulative SBF sample dependence on systemic velocity, with a breakdown by source.


\section{Distances to the Galaxy Hosts of SNe}
\label{sec:sn}

There are two predominant types of supernovae: those associated with stars with degenerate cores (type Ia) and those associated with stellar core collapse (type II).  The former are well established as providing a methodology for measuring distances with high accuracy.  The latter has seen less use and have reduced accuracy per event.  We focus on samples of both.


\subsection{Distances to the Galaxy Hosts of SNe Ia}
\label{sec:sn1a}

SNe~Ia arise in binary systems where at least one of the stars is a white dwarf.  The path, or possibly paths, resulting in explosions is not resolved, with the two most discussed alternatives being, alternatively, single-degenerate models where matter from a companion accretes onto a white dwarf, or double degenerate models where two white dwarfs merge \citep{2014ARA&A..52..107M}.  In spite of this uncertainty in the physical mechanism,
the minimal dispersion in the absolute luminosities of SNe~Ia after a calibration based on post-maximum decline rate \citep{1993ApJ...413L.105P} can be exploited to measure galaxy distances with high accuracy.  The events are sufficiently bright that SN Ia at $z<0.1$ can be discovered and monitored with moderate-sized telescopes.  This methodology can provide distances with 2-3 times the accuracy of TF and FP with considerably greater reach.  SNe are serendipitous events, though, and the number of well-studied occurrences remains small.   

Modest samples of SN Ia were included in earlier releases: 308 cases in {\it Cosmicflows-2}, increased to 389 in {\it Cosmicflows-3}.  Here, those samples are augmented by contributions from four new sources, more than doubling the available sample to 1008 SN Ia events within $z=0.1$.  Core-collapse SN II are proving to be useful distance tools also, although not with the same accuracy. A 
limited sample of SNII will be discussed in \S \ref{sec:snii}.

The new samples beyond those of {\it Cosmicflows-3} include 235 hosts from the Lick Observatory Supernova Search \citep{2013MNRAS.433.2240G}, 137 hosts from the first release of the Carnegie Supernova Project \citep{2018ApJ...869...56B}, 669 hosts from the collection by \citet{2021MNRAS.tmp.1412S}, 597 hosts in the PantheonPlus compilation \citep{2022ApJ...938..110B, 2022ApJ...938..113S}, the 560 late-type hosts in the largely overlapping and augmented Pantheon+SH0ES sample \citep{2022ApJ...934L...7R}, 134 hosts of SN~Ia with spectral {\it twin} properties by the Nearby Supernova Factory \citep{2021ApJ...912...71B}, and 89 hosts of events studied at infrared bands \citep{2019ApJ...887..106A}.  

There are considerable overlaps between these and the earlier contributions, as is evident if the contributions are summed.  Diverse light curve fitters are involved.  Selection cuts differ between programs and analyses could involve different photometry in different passbands but it is reassuring that alternate distance measurements to the same events are in agreement.  For 200 pairwise comparisons of multiple distance entries to the same SN Ia the weighted average difference of 0.063 mag in the modulus was found, about $3\%$ in distance.  The agreement is considerably less good in the distance estimates to multiple SN Ia in the same host.  Twenty-one hosts have measurements for at least two SN Ia events, with three such events in NGC\,3417 and four in NGC\,1316 in the Fornax Cluster.  To avoid issues related to differences in analysis procedures and zero-point scaling we consider just the 13 hosts with multiple events in the SH0ES study \citep{2022ApJ...934L...7R}.  The rms scatter of 0.278 mag corresponds to an uncertainty of $13.6\%$ in distance.  This large scatter is puzzling, especially since peculiar velocities are removed as a factor with events in a common host.  By comparison, \citet{2019ApJ...887..106A} and \citet{2021ApJ...912...71B} contend that the scatter they find with conventional light-curve fitters at optical bands correspond to $\sim 7\%$ in distance, with improvements to $5\%$ with the inclusion of infrared photometry or spectral matching ({\it twins embedding}) respectively.

\begin{figure}[]
\centering
\includegraphics[width=0.51\textwidth]{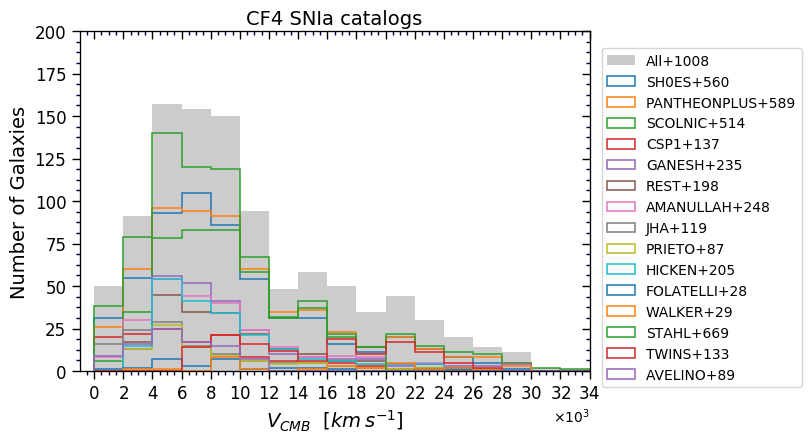}
\caption{
Cumulative histogram of SN Ia targets with systemic velocity and a breakdown by subsample as given by the legend. \label{fig:histv_SNIa}
}
\end{figure}

\begin{figure*}[!]
\centering
\includegraphics[width=1.05\linewidth]{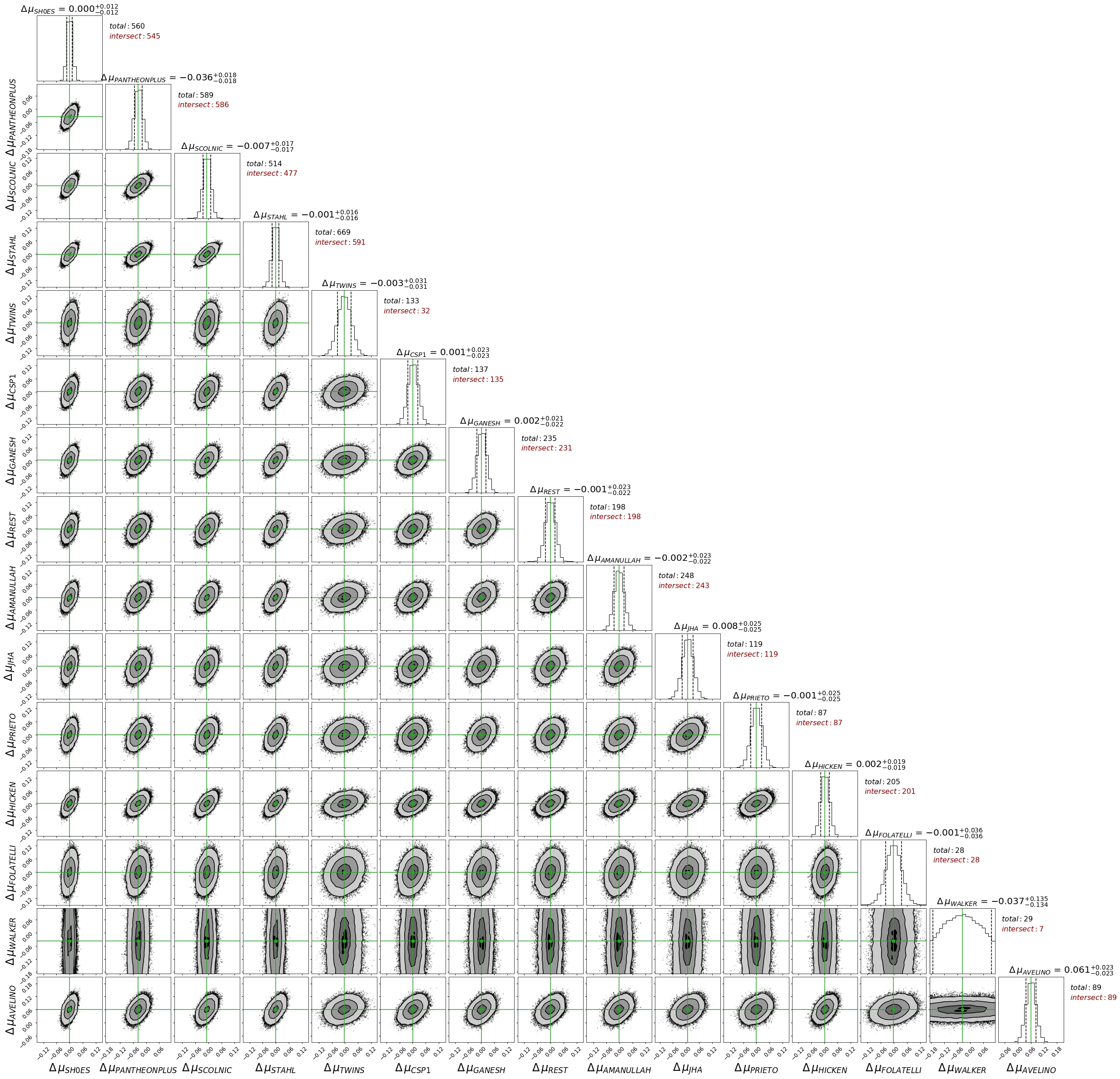}
\caption{
The posterior distribution of the optimized zero-points of SN Ia catalogs with respect to the SHOES distance moduli. 
Each panel covers $\pm0.15$ mag about the center of the distribution. Other details are as in Figure \ref{fig:mcmc_tf}. 
\label{fig:mcmc_SNIa}
}
\end{figure*}

\begin{figure*}[!]
\centering
\includegraphics[width=0.99\textwidth]{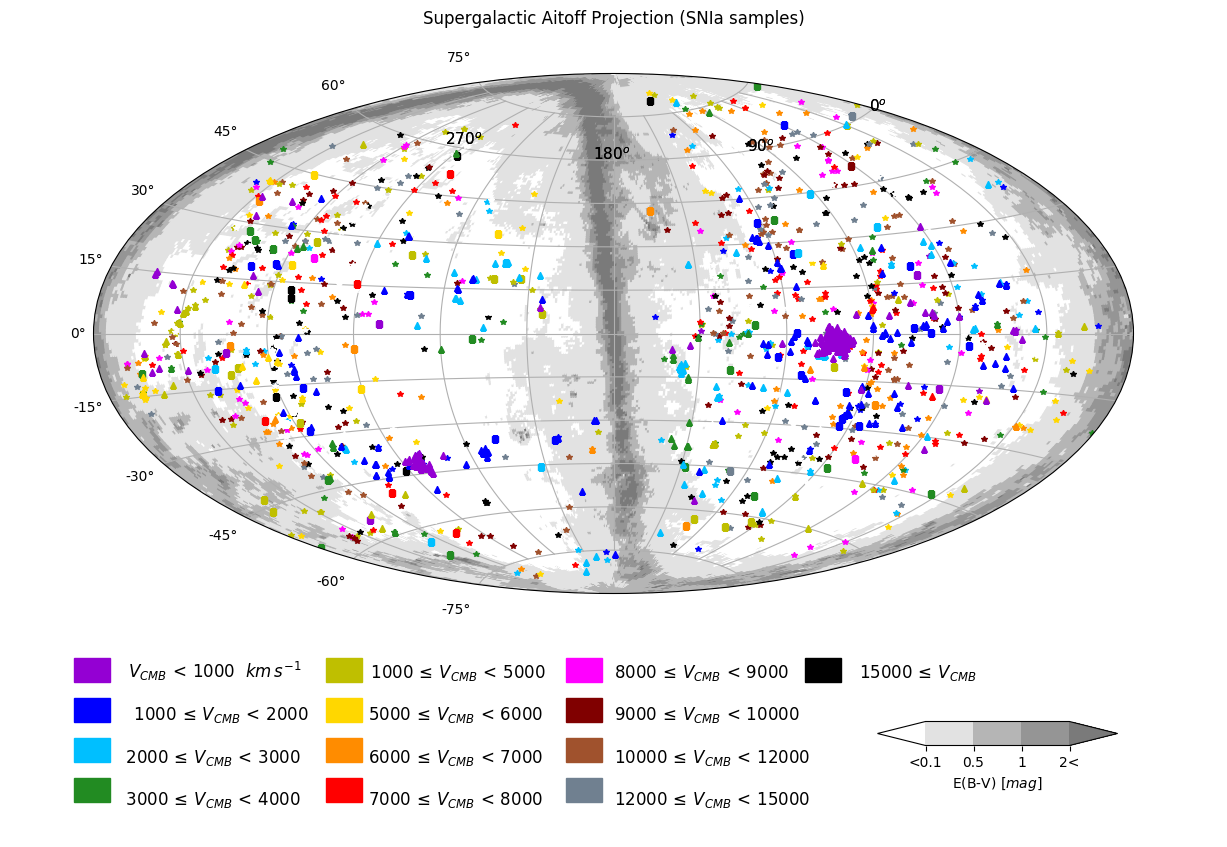}
\caption{
An aitoff projection in supergalactic coordinates of the distribution of the ensemble of 1008 SN Ia (asterisks), 94 SN II (squares), and 480 SBF (triangles) galaxies.  Colors relate to systemic velocities of the group of a galaxy as given in the table below the map.  Milky Way extinction levels are cast in shades of gray.   
There are concentrations of SBF targets in the Virgo Cluster at the right center and the Fornax Cluster at the lower left.
\label{fig:aitoff_sn}
}
\end{figure*}

A histogram of the distributions in velocities of the SN Ia subsamples is shown in Figure~\ref{fig:histv_SNIa}. 
The 15 SN Ia samples are merged in a manner analogous to that discussed in \S \ref{sec:bayesianMerge}.  The corner plot of moduli offsets following from the MCMC analysis is seen in Figure~\ref{fig:mcmc_SNIa}. The zero-point scaling is set to that provided by the SH0ES sample although, as with all the individual methodologies, the absolute scaling is subject to revision with the merging of all available material.   Figure~\ref{fig:aitoff_sn} illustrates the distribution of SN Ia on the sky along with the SN II and SBF contributions.  Undersampling in the celestial south accounts for the underrepresentation of events at supergalactic longitudes $180^{\circ}-270^{\circ}$ to the left of the zone of obscuration in the plot.  Concentrations of SBF contributions can be seen at the positions of the Virgo and Fornax clusters, at the right central and lower left, respectively. 

\subsection{Core collapse SNe II}
\label{sec:snii}

Although SN II vary intrinsically by more than two magnitudes, the underlying events are better understood than is the case with SN Ia and their luminosities correlate with observable properties.  More luminous SN~II have higher photospheric expansion velocities \citep{2002ApJ...566L..63H} and are bluer.  Colors are a monitor of extinction.  In {\it Cosmicflows-4} we include a sample of 96 SN II in 94 hosts studied by \citet{2020MNRAS.495.4860D, 2020MNRAS.496.3402D}.  The $1\sigma$ scatter about the Hubble diagram is $\sim 15\%$ in distance, about a factor 2 worse than realized with SN Ia.


\section{TRGB}
\label{sec:trgb}

The TRGB method for acquiring distances has two particularly resonant impacts on our program.  Relative accuracies (not accounting for zero-point issues) are typically at the level of 5\% with HST observations.  On the one hand, given an absolute calibration, the numerous measurements now available with overlaps with other methodologies play an important role in setting the absolute scale for extragalactic distances \citep{2019ApJ...882...34F, 2022ApJ...932...15A}.  On the other hand, the dense coverage of sources locally affords unprecedented information on a multitude of research interests.  The current edition of {\it Cosmicflows} provides TRGB distances to 489 galaxies \citep{2021AJ....162...80A}.

TRGB and the SBF procedure discussed in \S \ref{sec:sbf} have closely related physical bases.  With SBF, the brightest stars on the red giant branch are unresolved and distances are inferred from the statistical properties of image mottling. With TRGB the individual brightest red giant stars can be isolated and their brightness measured.  There is extensive experience with the use of the TRGB as a distance tool \citep{1990AJ....100..162D, 1993ApJ...417..553L, 1995AJ....109.1645M, 2002AJ....124..213M, 2006AJ....132.2729M, 2007ApJ...661..815R, 2014AJ....148....7W, 2017AJ....154...51M, 2017ApJ...835...28J, 2018SSRv..214..113B}. 

In practice, SBF and TRGB are most effective in two distinct regimes.  There are concerns with contamination from young populations and metallicity effects that are addressed in different ways.  In the case of SBF, the most stable candles are high surface brightness, very old, and high metallicity systems, with relative intrinsic magnitude constancy in the infrared.  In the case of TRGB, the optimal target stars have low metallicity and are observed near to $900~\mu$m where metallicity and age effects are minimal.  Such stars are best sought in the halos of galaxies where there is minimal extinction or crowding or contamination from young populations.

While several groups have been involved with TRGB studies, with a partial literature given above, all results reported in {\it Cosmicflows} have been drawn from data in the HST archives and analyzed in a uniform way with our standard procedures \citep{2009AJ....138..332J, 2021AJ....162...80A}.  Proceeding from {\it Cosmicflows-2} to $-3$ to $-4$, our collection of TRGB distances has grown from 297 to 384 to 489.  Ninety percent of galaxies brighter than $M_B = -13$ within 10 Mpc now have distance estimates with 5\% accuracy.  This material has given rise to population studies of individual systems \citep{2017MNRAS.464.2281M, 2017MNRAS.469L.113K}, to studies of the grouping properties of galaxies \citep{2017ApJ...843...16K}, and to studies of local filamentary structure and motions \citep{2017ApJ...850..207S, 2019ApJ...880...52A}.

\begin{figure}[]
\centering
\includegraphics[width=0.48\textwidth]{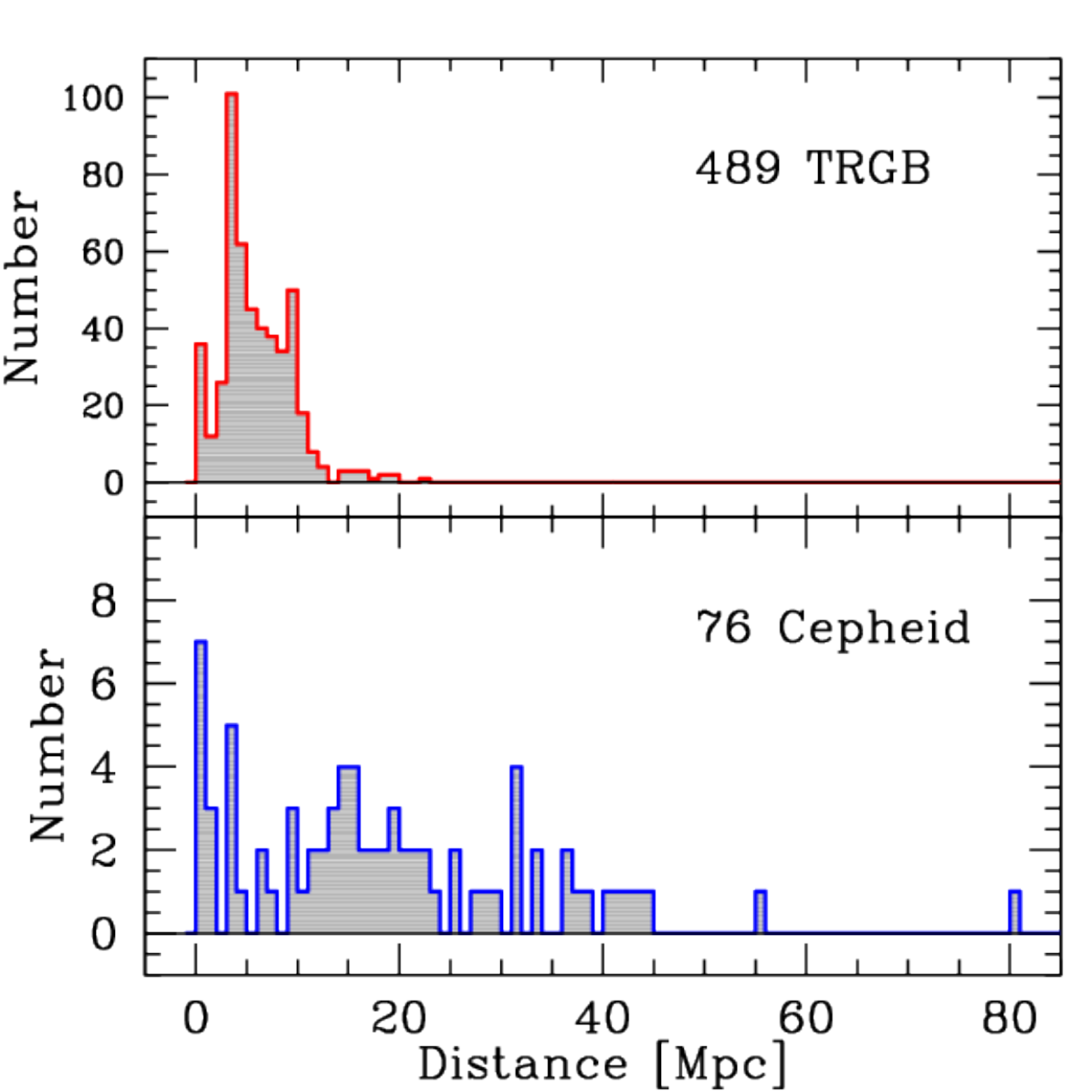}
\caption{Distribution in the distance of the TRGB and Cepheid period-luminosity samples.
 \label{fig:hist_trgb_ceph}
}
\end{figure}

The histogram of TRGB candidates with systemic velocity can be seen in Figure~\ref{fig:hist_trgb_ceph}.  Given that most of the TRGB systems are within 10~Mpc because of the single orbit capabilities of HST, these candidates are mostly to be found in the restricted sector of the sky of our home Local Sheet \citep{2008ApJ...676..184T}.  

The zero-point of our TRGB distances had been established in our previous publications with the study by \citet{2007ApJ...661..815R}.  That calibration followed from the identification of the HB in five Local Group galaxies (Sculptor, Fornax, and NGC~185 spheroidals, the irregular IC~1613, and the spiral M33) assuming the HB absolute magnitude found by \citet{2000ApJ...533..215C} and the implications for the absolute magnitude of the TRGB drawn from those systems.  More recent calibrations of the TRGB zero-point have led to some dispute \citep{2017ApJ...835...28J, 2017AJ....154...51M, 2019ApJ...882...34F, 2019ApJ...886...61Y, 2020ApJ...891...57F}.  It is a specific concern that the stellar trigonometric parallax underpinning of HB absolute magnitudes, and hence the entire Population~II distance ladder, is a product of Hipparcos satellite observations from the previous millennium.  Information from Gaia satellite is emerging that indicates that a zero-point rescaling is in order.  Attention will be given to this matter in \S \ref{sec:zp}.



\section{CPLR}
\label{sec:cplr}

The CPLR or Leavitt law \citep{1912HarCi.173....1L} has been considered the gold standard as a method for measuring absolute galaxy distances.  We accept a combined sample of 76 systems into {\it Cosmicflows-4}.  Our reference compilation of 40 objects is that given by \citet{2016ApJ...826...56R, 2019ApJ...876...85R, 2022ApJ...934L...7R}, see also \citet{2020ApJ...902...26Y,2021ApJ...913....3Y}, assuming the detached eclipsing binary distance modulus of $\mu_{LMC} = 18.477\pm0.004$ (stat) $\pm0.026$ (sys) of \citet{2019Natur.567..200P} ($d_{LMC} = 49.59\pm0.09$~Mpc).  We also use 23 nonoverlapping systems observed over the course of the HST key project to measure $H_0$ \citep{2001ApJ...553...47F} shifted to the LMC distance given above.  We include 12 systems all within $\sim 3$~Mpc observed from ground-based observatories and reported by \citet{2016AJ....151...88B} and the Araucaria Collaboration \citep{2017ApJ...847...88Z}, with scales set to our LMC distance.  The final case observed with HST and on our LMC scale is NGC~6814 \citep{2019ApJ...885..161B}.  A histogram of the measured distances of CPLR targets is shown in Figure~\ref{fig:hist_trgb_ceph}.

The absolute calibration of the CPLR remains a work in progress.  Cepheid variables are young stars that tend to live in crowded, dusty environments.  It remains to be resolved if there are metallicity dependencies.  There are expectations of a more refined scaling once parallax information provided by Gaia is digested \citep{2021ApJ...908L...6R}.  Our final scaling with {\it Cosmicflows-4} will be reviewed in \S \ref{sec:all}.


\section{Nuclear Masers}
\label{sec:maser}

The fortuitous occurrence of a water maser in the nucleus of NGC~4258 is making an out-sized contribution to studies of the extragalactic distance scale because of the well-defined geometry of the event \citep{2013ApJ...775...13H, 2019ApJ...886L..27R}.  Observations of the positions, velocities, and accelerations of the maser signals can be modeled to provide a direct geometric estimate of the angular-diameter distance of the galaxy without recourse to information external to the system.  The situation in the case of NGC~4258 is particularly favorable because the plane of rotation of the masers is almost edge-on and the very long baseline radio interferometric signals are strong.  The measured distance modulus is $29.397\pm0.032$, a formal error of only 1.5\% in distance. 

The occurrence in NGC~4258 is fortuitous indeed because extensive searches have, to date, only turned up five more cases that have warranted the attention leading to publication \citep{2020ApJ...891L...1P}.  Uncertainties vary wildly (median $\sim10\%$) because signals are weak and geometries are less favorable.  Distances for all of these systems are included in {\it Cosmicflows-4} at their geometrically determined values.  While nongeometric distances are subject to modification in the Bayesian analysis to follow, the maser distances (with associated uncertainties) are fixed.


\section{All Together}
\label{sec:all}

In the sections above, we have discussed the MCMC merging of subsamples within discrete methodologies.  In these cases, the analysis of targets in common has been made in separate studies.  Perhaps there have been independent observations, or distinct analysis procedures, but the premise is that there should be consistency in distance measurements by the same technique.  Scatter in multiple measures to the same galaxy should be (and are found to be) modest compared with the characteristic uncertainties in a method.

By contrast, distance information to a given galaxy by multiple methodologies is too rare to provide robust binding of the ensemble.  For example, if among the early-type FP and a late-type TF samples there is a galaxy in common (there are 67 cases among a total of 55,000 possible), then the result from at least one of the two contributions is suspect, given probable type confusion.  Then the overlaps involving SBF and SNe of either type Ia or II are modest because these samples are modest.  For the merging of these methodologies, we must look to coincidences in group affiliations.

The 55,681 galaxies with distances by at least one of these five sources find their way into 37,838 groups, including groups of one.  Some of the groups are well represented.  For example, in the case of the Coma Cluster (Abell\,1656; 1PGC\,44715), there are 209 FP measures, 50 TF measures, seven SN Ia hosts and two SN II hosts.  In the adjacent Leo Cluster (Abell\,1367, 1PGC\,36487) the numbers are 66, 49, 2, and 1, respectively.  In the more distant Hercules Cluster (Abell\,2147/51, 1PGC\,56962) there are 193 FP contributions, 60 by TF, and two SN Ia.  In the nearby Virgo Cluster (1PGC\,41220) we find 32 FP, 49 TF, an impressive 132 SBF, and four SN Ia.  Overall, there are 694 groups with FP$-$TF intersections.  SBF has 67 intersections with FP and 142 intersections with TF, cumulatively in 161 groups.

At this juncture, TRGB, CPLR, and maser contributions have not been mentioned.  These elements require special care because the targets are close by and the measurements are accurate enough to resolve their constituent groups.  Indeed, attention needs to be paid to this concern with SBF and SN Ia targets within $\sim2000$~\kms.


\subsection{Merging TF, FP, SBF, SN Ia, and SN II}
\label{sec:merge5}

\begin{figure*}[!]
\centering
\includegraphics[width=0.9\linewidth]{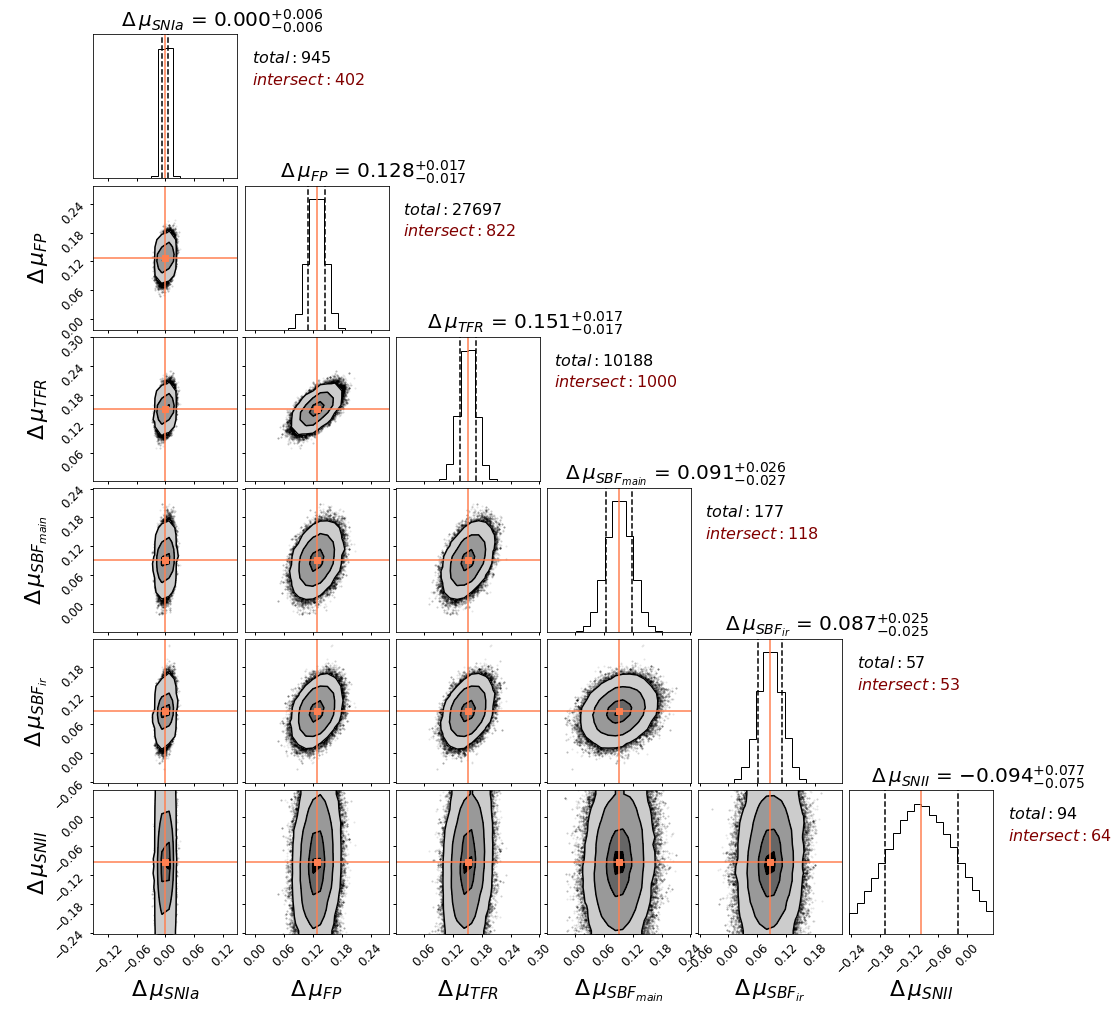}
\caption{
The posterior distribution of the optimized zero-points of the group catalogs with respect to the SN Ia catalog. 
Each panel covers $\pm0.15$ mag about the center of the distribution. Other details are as in Figure \ref{fig:mcmc_tf}. 
\label{fig:mcmc_groups}
}
\end{figure*}

Our procedure to achieve the union of the five methodologies, TF, FP, SBF, SN Ia, and SN II, is conceptually the same as that described in \S \ref{sec:bayesianMerge} with application to each of the individual methodologies.  The fundamental difference is that, whereas with the individual methodologies the bonding involves matching between galaxies, now the linkage units are galaxy groups.  So, for example, the 42,254 galaxies with FP distances are bundled into 27,701 groups.  The distance to a group is the weighted average of the moduli of all the FP measures in the group, with a statistical error given by the inverse square root of the sum of the weights.\footnote{Individual weights are formed from the inverse square of rms uncertainties.}  The velocity of the group is found by averaging over all known group members (not just those with distance estimates).  In the case of TF, the 12,395 individual galaxies with distance estimates are bundled into 10,189 groups.  With SBF, 425 individual galaxies lie in 227 groups.  The 1008 individual SN Ia hosts are found in 945 groups while the 94 SN~II hosts are all in 94 separate groups.

As with the previous MCMC analyses of individual methodologies, one of the contributions is selected as a zero-point reference. Here, we take the grouped SN~Ia sample as that reference.  This choice is impelled by three factors.  For one, the SN Ia sample provides the best cross-reference overlaps with the numerically dominant FP and TF components. Also, there is utility in the greater accuracy of SN Ia distances.  Finally, it is most useful that the SN~Ia are widely dispersed and have great reach in distance.  
At the onset of the joint analysis, each methodology had been set on its own arbitrary zero point scale.  For example, the SN Ia scale is consistent internally with H$_0\sim71$~\kmsMpc, the TF scale is consistent with H$_0\sim75$~\kmsMpc, and FP, SBF, and SN II are intermediate.
Now, all the methods will be brought to the same, but still arbitrary, zero-point scaling within statistical uncertainties.

The corner plot resulting from the MCMC merging of methodologies through joint group affiliations is shown in Figure~\ref{fig:mcmc_groups}.  The constraints on the unions with FP and TF are robust because the intersections are numerous and can involve groups with many distance measurements.  The constraints with SBF are reasonably good although intersections are limited because individual SBF measures have high accuracy.  The constraints are weakest in the connection with the currently limited SN II methodology.


\subsection{Zero-point Calibration}
\label{sec:zp}

The five methodologies that reach substantial distances have been merged to give coherent distances but on an arbitrary scale.  It is through a matching of sources with TRGB or CPLR measurements that the ensemble is put on an absolute scale, which manifests in an estimate of the Hubble constant.  Each of TRGB and CPLR are themselves anchored by other geometric techniques.  Both take grounding in the nuclear maser distance of 7.576~Mpc to NGC\,4258, which has a quoted uncertainty of $1.5\%$ \citep{2019ApJ...886L..27R}.  A separate foundation for CPLR is the quoted $1\%$ eclipsing binary distance of 49.59~kpc to the Large Magellanic Cloud \citep{2019Natur.567..200P}, the host galaxy for the largest coherent sample of Cepheid variables.  Both CPLR and TRGB take recourse in parallax measurements of appropriate stellar populations within the Milky Way.  In the case of CPLR, eight measurements come from HST WFC3 spatial scanning and 75  parallaxes are available from Gaia EDR3 \citep{2018ApJ...861..126R, 2021ApJ...908L...6R}. In the case of TRGB, the parallax targets are abundant Population~II stars that provide specification of the Horizontal Branch \citep{2000ApJ...533..215C}.  \citet{2007ApJ...661..815R} identified the levels of the Horizontal Branch in five Local Group galaxies observed with HST (Fornax, Sculptor, and NGC\,185 dwarfs, IC\,1613 and M33) to determine color-dependent values for the absolute magnitude of the TRGB.

The \citet{2007ApJ...661..815R} TRGB scale can be compared with TRGB on the scale in \citet{2019ApJ...882...34F}.  For 12 galaxies, there is the difference $<\mu_{rizzi} - \mu_{freedman}> = -0.028\pm0.035$ with rms scatter $\pm0.052$ \citep{2022ApJ...932...15A}.  This $1.5\%$ difference in distances is not statistically significant but the Rizzi scale distances closer are consistent with the \citet{2022ApJ...932...15A} determination of a Hubble constant value $0.9\%$ higher than that by \citet{2019ApJ...882...34F}.  

There are indications that the Population~II zero-point scaling will be altered somewhat as
Gaia satellite Data Release 3 (DR3) geometric parallaxes replace older Hipparcos parallaxes.  The following presents a best-effort estimate of the Gaia informed rescaling in advance of access to Gaia DR3.  We begin by recalling that the basis for the \citet{2007ApJ...661..815R} calibration was the following formula for the luminosities of stars on the Horizontal Branch (HB) derived from Hipparcos satellite trigonometric parallaxes \citep{2000ApJ...533..215C}
\begin{equation}
    M_V(HB) = 0.13 ([Fe/H] + 1.5) +0.54
    \label{eq:hb}
\end{equation}
The same reference gave a related formula for the luminosities of the RR Lyrae stars (RR) that lie embedded in the HB.
\begin{equation}
    M_V(RR) = 0.18 ([Fe/H] + 1.5) +0.5
    \label{eq:rr}
\end{equation}
Applied to the Large Magellanic Cloud, this reference finds $\mu_{LMC} = 18.54\pm0.03\pm0.06$ (statistical and systematic errors, respectively).  An updated version of Eq.\ref{eq:rr} based on parallaxes from the early release of Gaia DR3 (eDR3) is \citep{2022MNRAS.513..788G}
\begin{equation}
    M_V(RR) = 0.33 ([Fe/H] + 1.5) +0.63
    \label{eq:rr22}
\end{equation}
and these authors find a distance to the LMC of $\mu_{LMC} = 18.501\pm0.018$.  The change in the magnitude coefficient at fiducial [Fe/H]$= -1.5$ implies RR Lyrae stars fainter by 0.06~mag (although note the greater metallicity dependence).  The application to the LMC would have the LMC closer by 0.04~mag.

Consider another study.  \citet{2022ApJ...932...19N} derive new luminosity-metallicity and period-luminosity-metallicity relations with luminosities in Wesenheit magnitude units to negate reddening effects for 36 RR Lyrae in the Milky Way with eDR3 parallaxes and for RR~Lyrae in 39 nearby dwarf galaxies, making a simultaneous Bayesian fit to all Milky Way anchors and all RR Lyrae stars in each dwarf.  They find an offset for the Milky Way calibration stars from values obtained by \citet{2000ApJ...533..215C} $\mu_{gaia} - \mu_{hipparcos} = -0.048$ (without giving an error).  The subsequent distances obtained for the nearby dwarfs \citep{2022ApJ...932...19N} can be compared with our TRGB distance estimates with the Rizzi scaling \citep{2021AJ....162...80A}.  In two of the cases, NGC\,147 and NGC\,185 with the RR Lyrae information from \citet{2017ApJ...842...60M}, the modulus discordances are almost 0.4~mag and the reported RR Lyrae distances do not make sense given the quality of the TRGB values.  Discounting these two cases, eight available comparisons give $<\mu_{rizzi} - \mu_{rrl}> = 0.085\pm0.034$ with rms scatter $\pm0.095$. 

Rather than bootstraps to TRGB luminosities through RR Lyrae or the Horizontal Branch, there is the prospect of direct measurements of the luminosities of the brightest stars on the red giant branch, including dependencies on metallicity, from Gaia parallax measurements of large numbers of stars within the Milky Way.  A preliminary example is provided by the study by \citet{2021ApJ...908L...5S} of the determination of the TRGB magnitude for the $\omega$ Centaurus globular cluster with Gaia eDR3 parallaxes.  Their determination of $M_{I,trgb} = -3.97\pm0.06$ is lower by $\sim 0.08$ than the \citet{2007ApJ...661..815R} expectation.

A common thread can be seen through the discussion in this section.  Distances following from the Hipparcos-based \citet{2007ApJ...661..815R} calibration are too great, although the amplitude of the problem is uncertain.  From \citet{2022MNRAS.513..788G}, the revised scaling of RR Lyrae magnitudes and corresponding distance to the LMC suggests a modification between 0.04 and 0.06 mag.  From \citet{2022ApJ...932...19N}, changes in distances to Milky Way RR Lyrae and nearby dwarfs with observed RR Lyrae indicate a modification between 0.048 and 0.085. From  \citet{2021ApJ...908L...5S}, there is the single direct TRGB measurement giving the modification 0.08.  In no individual case is the justification for a modification compelling, yet the hints are strong.  Gaia DR3 observations are expected to provide clarity.  Under the circumstances and as a provisional solution, we will force agreement of our TRGB measurement for the galaxy NGC\,4258 \citep{2022ApJ...932...15A} to match the maser distance to that galaxy \citep{2019ApJ...886L..27R} rounded to the nearest $100^{th}$ of a magnitude, a zero-point shift to smaller distances with respect to \citet{2007ApJ...661..815R} of 0.05 mag.  The weakly color-dependent TRGB calibration becomes
\begin{equation}
    M_{I,trgb} = -4.00 + 0.22 [(V-I) - 1.6]
    \label{eq:Mtrgb}
\end{equation}
The uncertainty in the lead coefficient is dominated by systematics between Hipparcos and Gaia parallaxes that remain to be resolved, but an estimate $\pm0.03$ brackets the range of shifts discussed above.  The color term is also subject to review.  As opposed to the linear ramp with color incorporated in Eq.\ref{eq:Mtrgb}, published alternatives are a constancy with color on the premise that the measurement is made of the most metal poor of TRGB stars \citep{2019ApJ...882...34F} or a curved refinement of our linear relation \citep{2017ApJ...835...28J}.

\begin{figure}[!]
\centering
\includegraphics[width=0.99\linewidth]{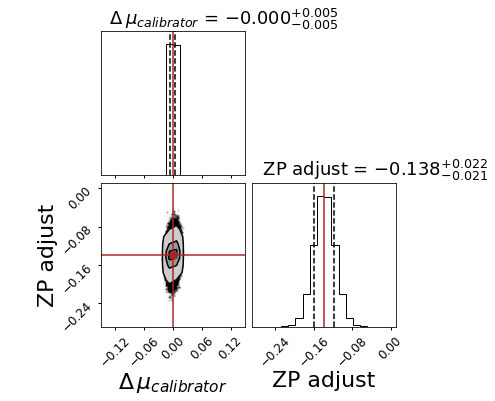}
\caption{
The posterior distribution of the zero-point adjustment.
Each panel covers $\pm0.15$ mag about the center of the distribution. Other details are as in Figure \ref{fig:mcmc_tf}. 
\label{fig:mcmc_zp_individuals}
}
\end{figure}

Direct comparisons can be made between CPLR distance moduli and TRGB moduli with the revised zero-point calibration using Gaia information.
 In the case of eight galaxies (including NGC\,4258) with TRGB estimates \citep{2021AJ....162...80A} and CPLR estimates from the SH0ES Collaboration \citep{2022ApJ...934L...7R} there is an average difference in distance modulus $<\mu_{trgb} - \mu_{cplr}> = 0.028\pm0.038$ with rms scatter $\pm0.107$.   There are 10 galaxies in common between the TRGB distances of \citet{2021AJ....162...80A} and CPLR distances of \citet{2001ApJ...553...47F} (assuming LMC is at $\mu_{LMC}=18.477$).  In two cases (M66 and NGC\,5253), the CPLR moduli are greater than TRGB by $\sim0.2$~mag.  It can be seen by eye with the TRGB fits displayed in the Extragalactic Distance Database \citep{2021AJ....162...80A} that a deviation this large is not supported by the TRGB measurements.  Considering the other eight cases, there is an average difference in modulus of $<\mu_{trgb} - \mu_{cplr}> = 0.019\pm0.025$ with rms scatter $\pm0.069$.  Combining the \citet{2022ApJ...934L...7R} and \citet{2001ApJ...553...47F} CPLR results, providing 16 comparisons, we find $<\mu_{trgb} - \mu_{cplr}> = 0.023\pm0.022$ with rms scatter $\pm0.087$.  There is agreement between the TRGB and CPLR observations in common to within $1\%$ with an rms scatter including errors in both inputs of $4\%$.

Going forward, then, our cumulative sample of zero-point calibrators will consist of 489 galaxies with TRGB distance estimates \citep{2021AJ....162...80A} shifted closer by 0.05 mag in the modulus, 76 galaxies with CPLR distances on the scale of \citet{2022ApJ...934L...7R}, and six maser distances \citep{2020ApJ...891L...1P} including the important case of NGC\,4258 \citep{2019ApJ...886L..27R}.  Accounting for overlaps, 542 zero-point calibrator galaxies are involved. In the union of all methodologies, the distance moduli of these calibrators in the ensemble will not shift.

The absolute distance scale calibration is established by evaluating linkages between individual calibrators and targets within the five methods ensemble discussed in \S \ref{sec:merge5}. 
  We find 128 individual galaxies among the TRGB, CPLR, and maser calibrators with matches in the combined FP, TF, SBF, SN Ia, SN II samples.  After rejection of three outliers (PGCs 5896, 44982, and 68535) and four Local Group galaxies, the remaining 121 matches are linked through an MCMC chain as summarized in the corner plot of Figure~\ref{fig:mcmc_zp_individuals}. 
  It follows that the arbitrary five source scale that was set with SN Ia requires the revision 
  \begin{equation}
  \mu_{zp} = \mu_{5so} - 0.138
  \pm0.022~\rm{mag}. 
  \label{eq:zp_offset}
  \end{equation}
There is significant variation in the coefficient of the revision if the five sources of inputs are considered separately.  Individually (with respect to the scale of Fig.~\ref{fig:mcmc_SNIa}) $\mu_{zp} - \mu_{xxx}$ equals $-0.082\pm0.057$ (TF: 81 cases), $-0.053\pm0.022$ (SN Ia: 41 cases), $-0.223\pm0.069$ (SBF: 29 cases), and $-0.142\pm0.128$ (SN II: eight cases).  There are only three overlaps with FP.  The result with SN~Ia alone is anomalous.  The anomaly is reduced somewhat in a comparison between groups containing calibrators and SN Ia: offset $-0.064\pm0.022$~mag with 44 SN~Ia.  
The difference from the five source value given in Eq.~\ref{eq:zp_offset} would have a 3.5\% effect on the Hubble Constant ($\Delta H_0 = -2.5$~\kmsMpc).  This difference in the establishment of an absolute scale calibration between referencing to SN Ia alone vs. the coupling of SN Ia to other methodologies
is essentially identical to that found in the construction of {\it Cosmicflows-3} \citep{2016AJ....152...50T}, see the discussion involving Fig.~12 in that paper.  Going forward, we retain the more robust zero-point scaling given by Eq.~\ref{eq:zp_offset} but we recognize the plausibility of systematic errors at least as large as the ambiguity raised by the SN Ia anomaly.



\subsection{Systematic Errors and the Path Forward}

Our results support the contention that the apparent differences in the cosmic expansion velocity today between direct measurements of distances and velocities and expectations of the standard $\Lambda$ cold dark matter model from conditions in the early universe are real, but the case is not yet compelling.  In our analysis the numbers of measurements are large so statistical uncertainties are very small.  However, the possibilities of systematic errors are a concern.

Issues can be separated between the establishment of the absolute distance scale from local observations and possible systematics in measurements well beyond the calibrators.  Consider first the local problem.  There is the expectation that, within a year or two, precision parallaxes provided by the Gaia experiment will provide precise intrinsic luminosities for the kinds of stars important for distance work: Cepheids, RR Lyrae, and horizontal branch stars, and stars at the tip of the red giant branch.  The samples should be numerous enough to unravel metallicity and age effects.  Present uncertainties in the absolute scaling are at the level of several percent.  Surely it is not at the 10\% level required to reconcile the Hubble constant controversy.  In any event, we can expect that this problem will soon be resolved at the 1\% level.

Presently, the FP and TF methodologies provide the large samples required for depth and wide field coverage of the sky.  However with only $20\%-25\%$ accuracy per target, uncertainties at redshifts $0.05c-0.1c$ are $3000-7500$~\kms.  Systematics that can arise from selection biases at the level of only a few percent can have several hundred kilometers per second effects. If a sample is free of systematics it should pass our tests of Hubble parameter constancy with redshift or other parameters, but such tests are not a guarantee against problems.

Going forward, prospects for building a large all-sky sample at substantial redshifts look best with SN Ia.  Surveillance at high cadence with moderate aperture telescopes suffice to detect the $\sim 10^3$ SNe Ia within $z=0.1$ that erupt each year.  With samples of many thousand, with photometry at multiple passbands and accompanying spectra, it should be possible to empirically calibrate variations in the properties of the explosion events, providing relative distances accurate to 5\%.  Vulnerability to systematics should be reduced roughly linearly with the improved statistical accuracy.

What remains is the coupling between the near-field calibration and the far-field mapping. The approach of coupling between Cepheids and SN~Ia may be reaching its limits.  Proximate SNIa are few and the requisite monitoring required to characterize distant Cepheids will challenge resources.  Prospects are better with TRGB, bright in the infrared and accessible to 40~Mpc or more with a single exposure sequence per target with JWST.  TRGB targets will couple equally to SN~Ia and SBF hosts.  JWST will facilitate the SBF methodology with 5\% distance accuracy out to $z \sim 0.07$.  An all Population II TRGB$-$SBF path to galaxy distances will complement the Population I CPLR$-$SN~Ia path and the availability of two independent methodologies covering the same distance range will provide a good test of systematics.


\section{Summary of the Distance inputs}
\label{sec:summarytable}

\begin{deluxetable*}{llrl}
\centering
\scriptsize
\tablecaption{Summary of all distance contributions}
\label{tab:contributions}
\tablehead{ \\
Method & Subsample & Number of Distances & Reference \\
}
\startdata
TF & & 12223 galaxies in 10188 groups \\
 & cf4          &  9323 & \citet{2022MNRAS.511.6160K} \\
 & spitzer      &  2202 & \citet{2014MNRAS.444..527S} \\
 & cf2          &  4014 & \citet{2013AJ....146...86T} \\
 & sfi          &  3938 & \citet{2007ApJS..172..599S} \\
 & 2mtf         &  1712 & \citet{2019MNRAS.487.2061H} \\
 & flat         &   546 & \citet{2018MNRAS.479.3373M} \\
\hline
FP & & 42223 galaxies in 27691 groups \\
 & sdss         & 34045 & \citet{2022MNRAS.515..953H} \\
 & 6dfgsv       &  7099 & \citet{Qin2018} \\
 & smac         &   689 & \citet{2001MNRAS.327..265H} \\
 & enear        &   447 & \citet{2002AJ....123.2990B} \\
 & efar         &   696 & \citet{2001MNRAS.321..277C} \\
\hline
SBF & & 480 galaxies in 227 groups \\
 & vir/for      &   134 & \citet{2010ApJ...724..657B} \\
 & tonry        &   297 & \citet{2001ApJ...546..681T} \\
 & cantiello    &    89 & \citet{2018ApJ...856..126C} \\
 & ir/hst       &    63 & \citet{2021ApJS..255...21J} \\
 \hline
SN Ia & & 1008 galaxies in 945 groups \\
 & sh0es        &   560 & \citet{2022ApJ...934L...7R} \\
 & pantheonplus &   589 & \citet{2021arXiv211203863S} \\
 & scolnic      &   514 & Early Pantheon compilation  \\
 & csp1         &   137 & \citet{2018ApJ...869...56B} \\
 & loss         &   235 & \citet{2013MNRAS.433.2240G} \\
 & rest         &   198 & \citet{2014ApJ...795...44R} \\
 & union2       &   248 & \citet{2010ApJ...716..712A} \\
 & jha          &   119 & \citet{2007ApJ...659..122J} \\
 & prieto       &    87 & \citet{2006ApJ...647..501P} \\
 & constitution &   205 & \citet{2009ApJ...700.1097H} \\
 & folatelli    &    28 & \citet{2010AJ....139..120F} \\
 & walker       &    29 & \citet{2015ApJS..219...13W} \\
 & stahl        &   669 & \citet{2021MNRAS.tmp.1412S} \\
 & twins        &   133 & \citet{2021ApJ...912...71B} \\
 & avelino      &    89 & \citet{2019ApJ...887..106A} \\
\hline
SN II & & 94 galaxies in 94 groups \\
 & de Jaeger    &    94 & \citet{2020MNRAS.495.4860D} \\
\hline
TRGB & & 489 galaxies \\
 & edd          &   489 & \citet{2021AJ....162...80A} \\
\hline
CPLR & & 76 galaxies \\
 & riess        &    40 & \citet{2022ApJ...934L...7R} \\
 & keyproject   &    23 & \citet{2001ApJ...553...47F} \\
 & nearby       &    12 & \citet{2016AJ....151...88B} \\
 & bentz        &     1 & \citet{2019ApJ...885..161B} \\
\hline
MASER & & 6 galaxies \\
 & pesce        &     5 & \citet{2020ApJ...891L...1P} \\
 & ngc4258      &     1 & \citet{2019ApJ...886L..27R} \\
\hline
\enddata
\end{deluxetable*}

Table~\ref{tab:contributions} is a compilation of statistics and references pertaining to all the methodologies and all the subsamples contributing to the study.


\section{Cosmicflows-4 Catalog}
\label{sec:catalog}

The {\it Cosmicflows-4} compendium of galaxy distances is presented in three tables, with a representative stub of the first several lines of each given in the Appendix and the full tables available online and at the Extragalactic Distance Database (\url{https://edd.ifa.hawaii.edu}).  

The line entries in Table~\ref{tab:individual_galaxies} provide information on 55,877 individual galaxies with information pertaining to their distances.  Individual distance moduli and uncertainties are given, where available, for each of the methodologies, SN~Ia, TF, FP, SBF, SN~II, TRGB, CPLR, MASER, and a weighted average value, all on the absolute scaling established by the combined TRGB, CPLR, and MASER calibrators.  The table also provides coordinates, velocities in the CMB reference frame, and \citet{2015AJ....149..171T} and \citet{2017A&A...602A.100T} group affiliations.

Table~\ref{tab:galaxy_groups} provides related information for 38,057 groups.  By order of preference, the group constructions draw upon the groups of \citet{2017ApJ...843...16K}, \citet{2015AJ....149..171T}, and \citet{2017A&A...602A.100T} in that order. If a galaxy is not within the samples of those groups, it can receive a group linkage through position and velocity coincidence with the virial properties of an established group. For each group with a distance estimate, the number and averaged values within each methodology are given as well as an overall averaged distance modulus value, on the absolute scale established by the TRGB, CPLR, and MASER calibrators.  In the stub of this table in the Appendix, the groups are sorted by the number of group targets available with the FP methodology. 

Table~\ref{tab:group_vpec} provides derivative information for the same 38,057 groups, including coordinates, velocities in different reference frames, peculiar velocities following from Eqs.~\ref{eq:vds}$-$\ref{eq:vramp} discussed in the next section, and values of the Hubble parameter $H_i = f_iV_{cmb,i}/d_i$ for each group.


\begin{figure}[!]
\centering
\includegraphics[width=1\linewidth]{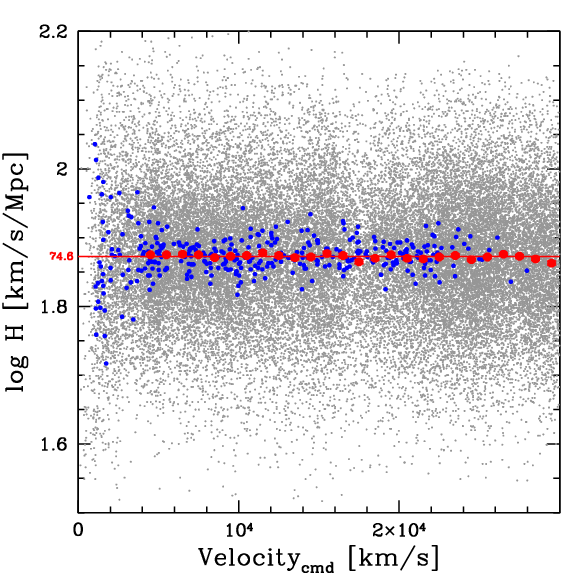}
\caption{
Hubble parameter vs. systemic velocity.  Individual groups are identified in gray while groups with uncertainties in modulus 0.10 mag or better are highlighted in blue.  Red symbols with (small) error bars denote mean log$H$ values in 1000~\kms\ intervals.  The overall average of values for $V>4000$~\kms\ is given by the horizontal red line.}
\label{fig:HV_all}
\end{figure}

\section{Properties of the Catalog}
\label{sec:properties}

\begin{figure*}[!]
\centering
\includegraphics[width=.48\linewidth]{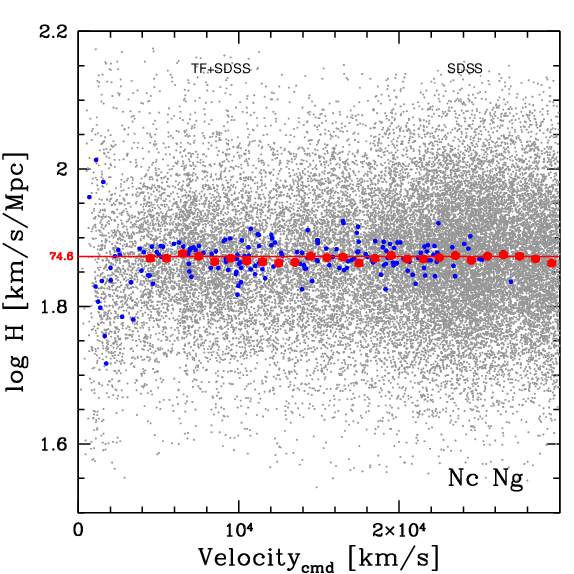}
\includegraphics[width=.48\linewidth]{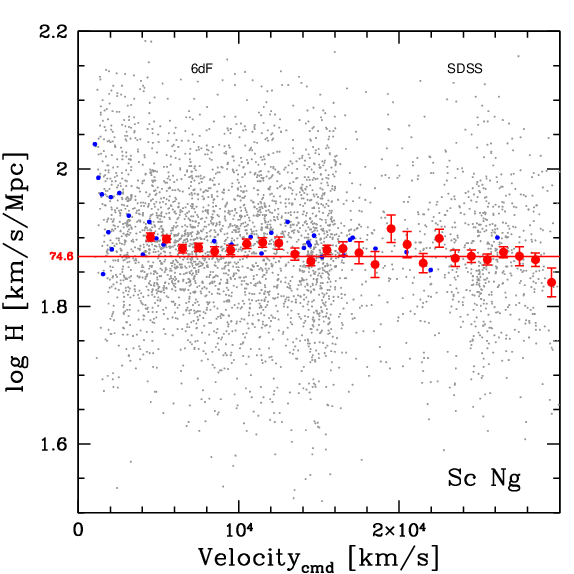}
\includegraphics[width=.48\linewidth]{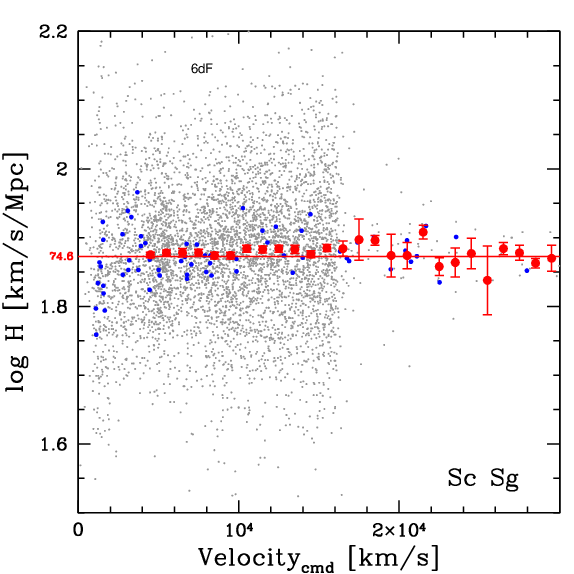}
\includegraphics[width=.48\linewidth]{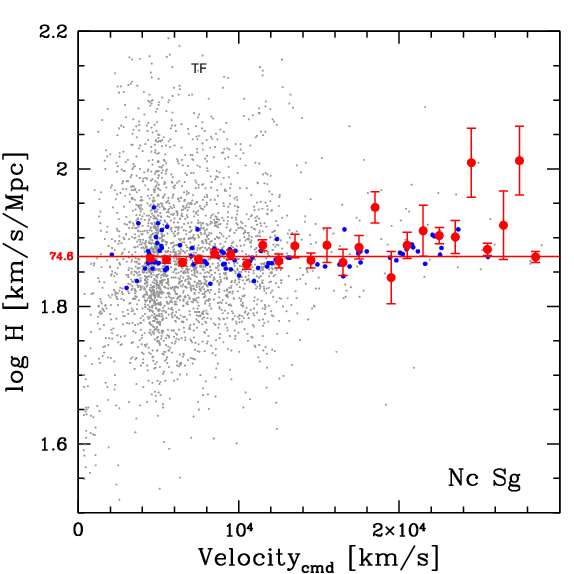}
\caption{
Hubble parameter vs. systemic velocity separated into four sectors: north (N) and south (S) celestial (c) and galactic (g) quadrants.  Gray, blue, and red symbols are the same as in Fig.~\ref{fig:HV_all}. Dominant contributing methodologies in each of the sectors are specified across the top of panels:  SDSS FP restricted to the north celestial and galactic cap, 6dFGSv FP restricted to the celestial south, TF everywhere across the sky but favoring the decl. range accessed by Arecibo Telescope, and other methods widely distributed but relatively sparse.  The means in log$H$ are noisy at $V>16,000$~\kms\ in sectors where the contributions come mostly from a few SN~Ia. Log$H$ values tend to lie above the mean line in the ScNg sector, in the direction toward the apex of the CMB dipole, and tend to lie below the mean line in the anti-apex NcSg quadrant.
}
\label{fig:HV_sectors}
\end{figure*}

With the expectation that peculiar velocities are of order 300~\kms, beyond about 4,000~\kms\ it is anticipated that motions due to cosmological expansion will dominate.  Accordingly, a necessary but not sufficient test our distance measures should pass is rough constancy in the value of the Hubble parameters with redshift after exclusion of the nearby realm.  Our collection of distances pass this test as seen in Figure~\ref{fig:HV_all}.  Here, each gray dot is at the Hubble parameter and velocity location of a group, including groups of one.  Heavier blue dots identify groups with uncertainties 5\% or better in distance.  
The overall average for groups at greater than 4,000~\kms\ is log$H_0 = 1.8725$ corresponding to $H_0=74.6$~\kmsMpc. The logarithmic rms scatter is $\pm 0.091$.  The statistical standard deviation on $H_0$ with 35,000 groups is only $\pm 0.2$~\kmsMpc, smaller than the uncertainty in the zero-point calibration through Eq.~\ref{eq:zp_offset} of $\pm0.8$~\kmsMpc.

Contributions to our collection of distances have distinct sky coverage.  There is the risk of offsets between contributions that would result in artificial flows between regions dominated by different inputs.  Our Bayesian linkages between methods minimize such problems. The expectation is that there would not be suspicious trends between sectors of the sky dominated by different sources.  The three numerically largest contributions come from the two FP programs; SDSS and 6dFGSv, which are restricted to the northern and southern sky, respectively, and the TF programs, which have an emphasis in northern sky coverage.  The displays in Figure~\ref{fig:HV_sectors} break down the all-sky coverage of Figure~\ref{fig:HV_all} into four sectors, whether north or south on the celestial sky and whether north or south of the galactic plane.  The sky coverage is obviously uneven.  The only abundant coverage beyond 16,000~\kms\ is in the sector north celestial and north galactic, as provided by SDSS FP.  A sprinkle of coverage at high velocities is provided by SN~Ia contributions.  Trends can be detected between sectors, most notably elevated Hubble values in the south celestial$-$north galactic sector and slightly depressed Hubble values in the opposite north celestial$-$south galactic sector.  This pattern can be attributed to the known motion toward the so-called great attractor \citep{1987ApJ...313L..37D} and the kinematic response associated  with the CMB temperature dipole \citep{1996ApJ...473..576F}.

\begin{figure}[!]
\centering
\includegraphics[width=.97\linewidth]{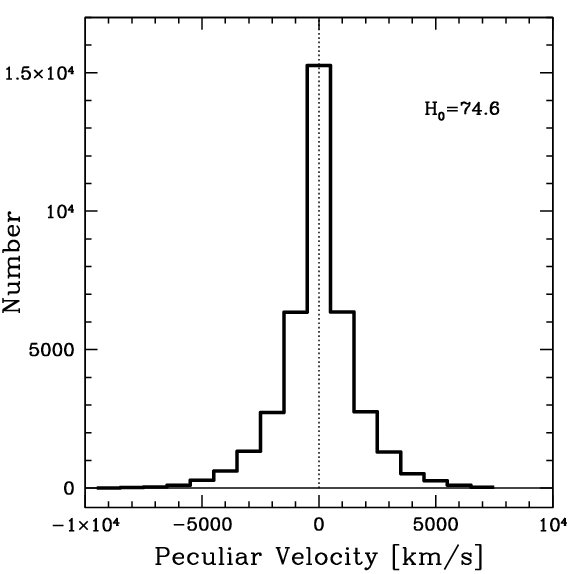}
\includegraphics[width=.97\linewidth]{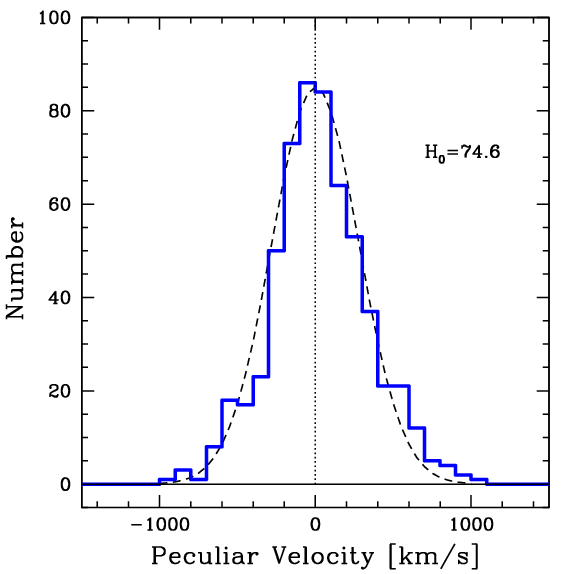}
\caption{
Top: Histogram of 38,057 group peculiar velocities. Bottom: histogram of peculiar velocities for 581 groups with distance uncertainties better than 5\%.  The dashed curve illustrates a Gaussian distribution with standard deviation of 280~\kms.}
\label{fig:hist_vpec}
\end{figure}

The observed radial motions of galaxies derive from a combination of the expansion of the universe and deviations assumed to arise from gravitational perturbations. This latter component, commonly called the peculiar velocity, can be extracted with knowledge of the distance of the system and of the mean expansion of the universe characterized by Hubble's constant, following 
\citet{2014MNRAS.442.1117D}
\begin{equation}
V_{pec}^{ds} = (fV_{cmb} - H_0 d)/(1 + H_0 d/c) .
\label{eq:vds}
\end{equation}
The cosmological curvature correction parameter $f$ was defined in connection with Eq.~\ref{eq:logh}.
As is well known, uncertainties in peculiar velocities skew to larger negative values in the translation from logarithmic distance modulus to distance.  A formulation to negate the skewness and dampen the amplitude of peculiar velocities in the presence of uncertainties was developed by \citet{2015MNRAS.450.1868W}
\begin{equation}
V_{pec}^{wf} = {{fV_{cmb}}\over{1+fV_{cmb}/c}} {\rm log}(fV_{cmb}/H_0 d) .
    \label{eq:vwf}
\end{equation}
This approximation has application if $v_{pec}<<H_0 d$.  For present rough purposes, we estimate peculiar velocities with Eq.~\ref{eq:vds} nearby where peculiar velocities can be a substantial fraction of observed velocities, transitioning to Eq.~\ref{eq:vwf} by velocity $V_{ls}=3000$~\kms\ in the Local Sheet frame \citep{2008ApJ...676..184T}, with the ramp
\begin{equation}
    V_{pec} = V_{pec}^{ds} (1-V_{ls}/3000) + V_{pec}^{wf} V_{ls}/3000 .
    \label{eq:vramp}
\end{equation}
A histogram of the peculiar velocities of all of the 38,000 groups, all those represented in gray in Figure~\ref{fig:HV_all}, is presented in the top panel of Figure~\ref{fig:hist_vpec}, while the bottom panel zooms to the histogram for groups with distance modulus uncertainties 0.10 mag or better, the blue points in Figure~\ref{fig:HV_all}.  The distribution of peculiar velocities with well-constrained distances is approximated by a Gaussian distribution with standard deviation 300~\kms.


\subsection{A preliminary look at sky coverage}
\label{sec:prelimlook}

\begin{figure}[!]
\centering
\includegraphics[width=.97\linewidth]{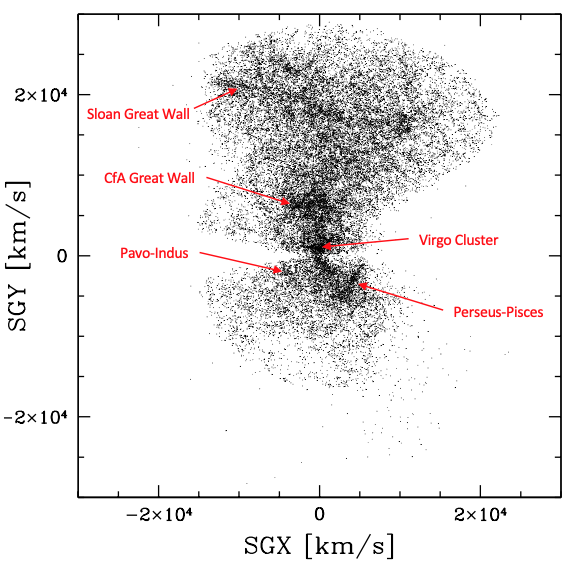}
\includegraphics[width=.97\linewidth]{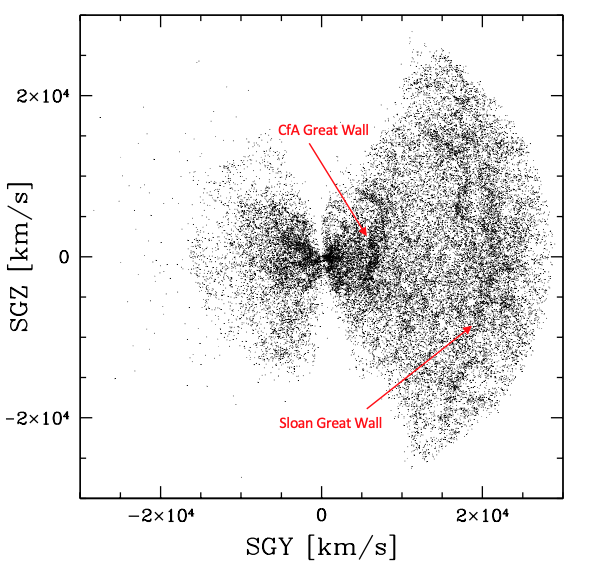}
\caption{
Two views of the ensemble of groups in supergalactic coordinates. The obscuration zone of the Milky Way lies slightly tilted at SGY$\sim 0$.
The extensive SDSS FP sample in the north galactic, north celestial sector is evident at SGY$> 0$.  Both the Sloan and CfA Great Walls are evident.  A sprinkle of points representing SN~Ia hosts can be seen within the domain extending to $z=0.1$.
}
\label{fig:2views}
\end{figure}


\begin{figure}[!]
\centering
\includegraphics[width=.97\linewidth]{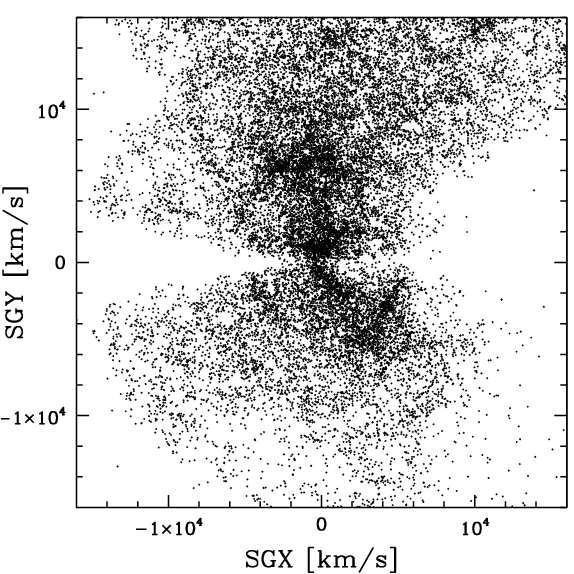}
\includegraphics[width=.97\linewidth]{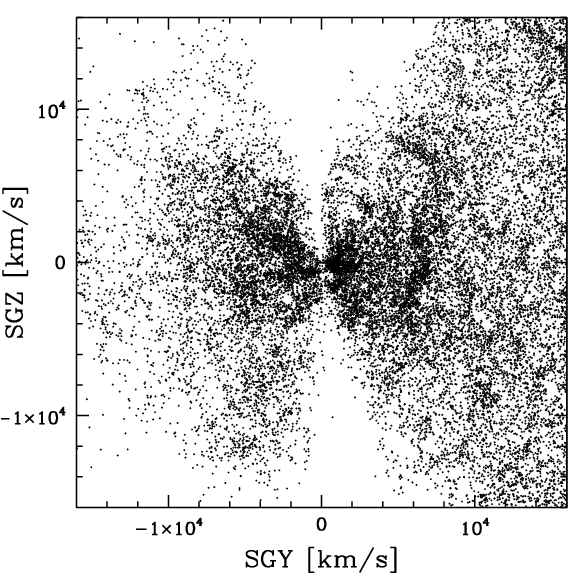}
\caption{
Zoom-in on the two panels of Fig.~\ref{fig:2views}.  Within 16,000~\kms\ there is reasonable sky coverage except at large positive SGX and the zone of obscuration. An \href{https://sketchfab.com/3d-models/the-cosmicflows-4-ensemble-of-groups-efb0aebb45b644c0ab4c5c5774452f75}{interactive 3D visualization} of the distribution of groups provides further insights.
}
\label{fig:2views_16}
\end{figure}

\begin{figure}[!]
\centering
\includegraphics[width=.97\linewidth]{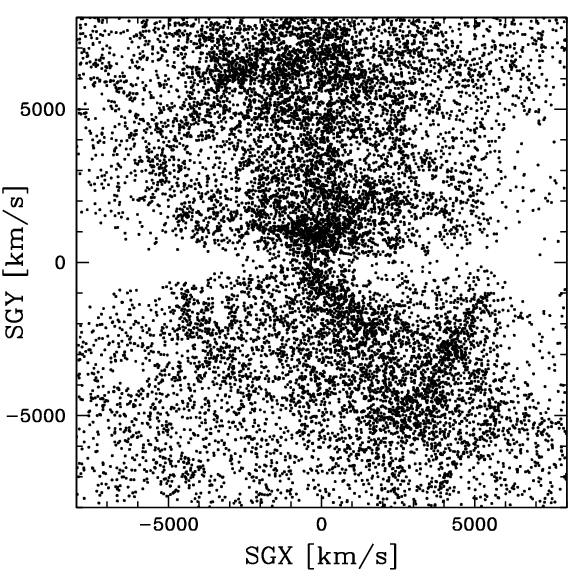}
\includegraphics[width=.97\linewidth]{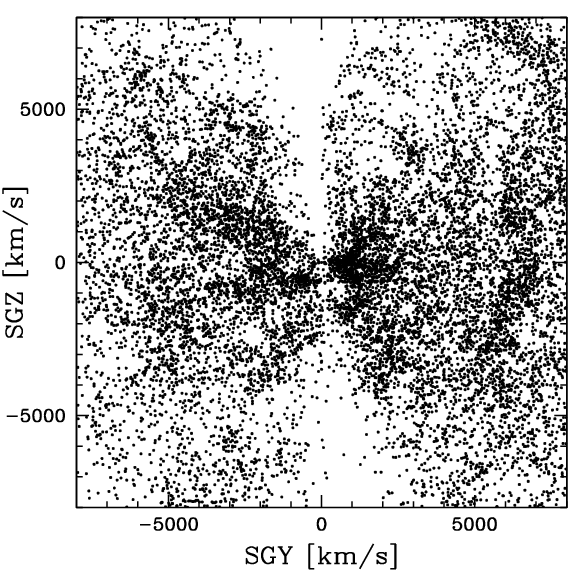}
\caption{
A further zoom-in on the two panels of Figs.~\ref{fig:2views} and \ref{fig:2views_16}.   Outside the zone of obscuration, there is reasonably uniform coverage with sky direction within this domain.
}
\label{fig:2views_8}
\end{figure}

\begin{figure}[!]
\centering
\includegraphics[width=.97\linewidth]{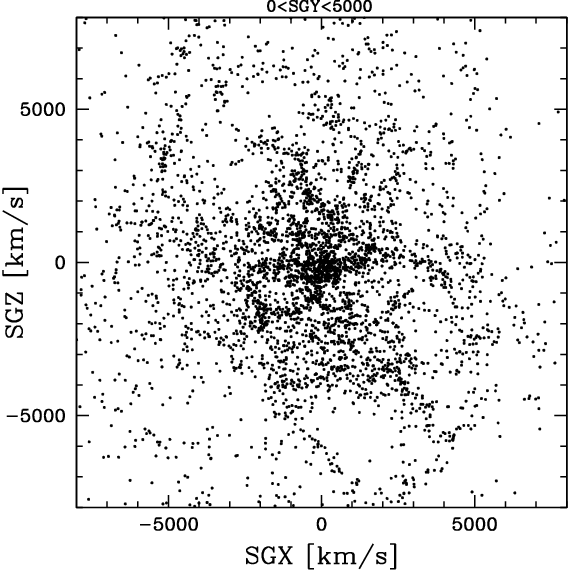}
\includegraphics[width=.97\linewidth]{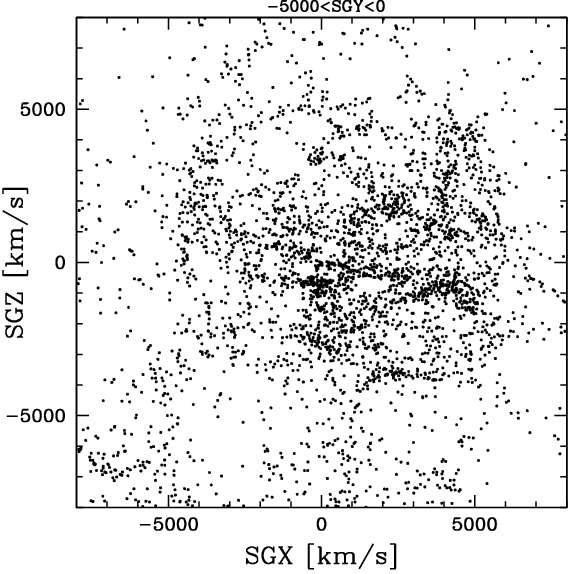}
\caption{
The view almost normal to the galactic plane restricted to $\pm5000$~\kms\ and separating between north of the Galaxy (top panel) and the galactic south (bottom).
}
\label{fig:2xzviews_8}
\end{figure}

The distribution of all the entities in the group catalog are shown in the two orthogonal projections of Figure~\ref{fig:2views}.  The coverage is dense and relatively uniform around the sky within $\sim 8,000$~\kms\ except for the Milky Way avoidance zone at SGY$\sim 0$.  The two FP samples provide coverage in two distinct fans: 6dFGSv out to 16,000~\kms\ in the celestial south and SDSS out to 30,000~\kms\ in the north galactic and celestial polar cap.  There is only spotty coverage in the sector in the northern sky south of the Galaxy at velocities greater than 8,000~\kms\ where the range of TF coverage falls off (see the histogram in Fig.~\ref{fig:histv_tf}).

The most dramatic feature seen in these projections contains as a component the Sloan Great Wall \citep{2004ogci.conf....5V, 2005ApJ...624..463G} at SGY$\sim 20,000$~\kms.  The Center for Astrophysics Great Wall \citep{1986ApJ...302L...1D} is seen a SGY$\sim 6,000$~\kms.

While there is vast terrain to explore opened up by the full SDSS FP sample, uncertainties in individual distance measurements are extreme in the outer regions, and there will continue to be particular interest in the volume of reasonably uniform coverage.  As can be seen in Figure~\ref{fig:2views_16}, the ensemble sample is reasonably dispersed within 16,000~\kms\ except for the falloff beyond $\sim8000$~\kms\ in the celestial north outside the boundaries of the SDSS study, at positive SGX and positive SGZ.  There is the usual caveat regarding the zone of obscuration.

Figures~\ref{fig:2views_8} and \ref{fig:2xzviews_8} expand views to the inner 8,000~\kms\ volume where the TF coverage is of principal importance and individual peculiar velocities are tractable.  Fig.~\ref{fig:2xzviews_8} shows 5,000~\kms\ slices in SGY on the north and south sides of the galactic plane in the direction orthogonal to the views of Figs.~\ref{fig:2views}$-$\ref{fig:2views_8}.  The complex networks of filaments await detailed morphological and dynamic studies.


\section{Summary}
\label{sec:summary}

{\it Cosmicflows-4} combines distances from eight methodologies: five reaching galaxies over a substantial volume and three providing anchors of the absolute scale.  The case for a controlled homogeneous approach to establishing distances is well founded \citep{2019ApJ...876...85R}, but the use of a heterogeneous sample also has merits.   The possibility of systematic errors is now more worrisome than statistical errors.  The exploration of multiple paths can reveal differences that warn of the occurrence of systematics.   

The various methodologies are meshed together by overlaps.  The only substantial overlaps between individual galaxies within {\it Cosmicflows-4} involve a moderate number of matches with SN~Ia events.  More robust linkages are through common memberships in groups.  Here, three group catalogs are employed, by order of priority those by \citet{2017ApJ...843...16K}, \citet{2015AJ....149..171T}, and \citet{2017A&A...602A.100T}.  The importance of the group linkages should not be underestimated.  In the extreme, in each of the clusters Coma and Hercules there are $\sim200$ FP measures and $50-60$ TF measures.  The two clusters have hosted nine of our SNe~Ia. 

Our absolute scaling also comes through group affiliations, those involving galaxies with TRGB, CPLR, or maser distances.  With this step, as with all the precursor steps involving the integration of individual methodologies and the uniting of methodologies, the merging was carried out with an MCMC sampling from probability distributions governed by error assignments.

Our best value for the Hubble constant of $H_0 = 74.6$~\kmsMpc\ has a formal statistical error of only $\pm 0.8$~\kmsMpc.  However, as discussed in \S \ref{sec:zp}, the calibration is lowered by 2.5~\kmsMpc\ if the only coupling is through SN~Ia.  Let us try to be clear on the separate routes to $H_0$ giving these different results.  With our preferred route, the 1008 SN~Ia are merged with the FP, TF, SBF, and SN~II material onto a common but arbitrary scale, then this ensemble is merged with 128 galaxies involving some combination of the TRGB, CPLR, and maser calibrators.  In the alternate route, the linkage involves establishing the SN~Ia scale alone with 44 SN~Ia host galaxies with TRGB, CPLR, or maser information.  This latter route giving $H_0=72.1$~\kmsMpc\ is in good agreement with the value by \citet{2022ApJ...934L...7R} of $H_0=73.0\pm1.0$~\kmsMpc\ involving much the same input.  The same systematic difference was found between an ensemble of methodologies approach versus SN~Ia only in the compilation of {\it Cosmicflows-3} \citep{2016AJ....152...50T}. 

Is the difference due to modest number statistics?  Or could the difference arise because SN~Ia involved in the zero-point calibration are systematically different from the ensemble of observed SN~Ia?  The absolute values of SN~Ia are suspected to vary with host properties, possibly related to the mean ages of explosions \citep{2013ApJ...764..191H, 2013ApJ...770..108C, 2018A&A...615A..68R}.  We are warned that systematics at the level of 3\% in the absolute scale are possible, not including the grounding by geometric parallaxes that Gaia DR3 should illuminate.   

In any event, the primary motivation for the {\it Cosmicflows} program is the study of {\it deviations} from cosmic expansion, which are independent of the absolute scale calibration.  Our goal is to compile a catalog of distances that, although they have uncertainties, are unbiased in relative distances.  There is no recourse to information about the distribution of mass through redshift surveys or other considerations.  It is well known that biases can arise in the inference of peculiar velocities from distances with large errors beyond what is captured by Eq.~\ref{eq:vwf}.  Mitigating efforts with past samples of galaxy distances involve Wiener filtering or Bayesian coupling with the inferred distribution of mass from redshift catalogs \citep{2012ApJ...744...43C, 2015MNRAS.449.4494H, 2021MNRAS.505.3380H, 2015MNRAS.450..317C, 2019A&A...625A..64J, 2019MNRAS.488.5438G, 2021MNRAS.507.1557L}.  It is expected that the availability of the present catalog of distances will lead to even more extensive studies.

There are deficiencies in the present catalog where improvements can be realized.  It is anticipated that the absolute scale foundation given by geometric parallaxes will soon be established more rigorously with the release of Gaia DR3.  Throughout, there is a need for greater overlaps between methodologies.  SN~Ia will continue to play an important role, with the possibility of accumulating samples of many thousands in the near future.  It will be difficult to seriously expand the CPLR samples with current space facilities.  However, the alternate Population II route of TRGB linked to SBF should be a profitable way forward with JWST.  There is the prospect of acquiring considerably larger samples than with CPLR$-$SN~Ia with comparable accuracy per target.  The availability of a second independent path will be a real test of systematics.


\section{acknowledgements}

We thank an anonymous referee for a gracious review.
Funding over the many years of the {\it Cosmicflows} project has been provided by the US National Science Foundation grant AST09-08846, the National Aeronautics and Space Administration grant NNX12AE70G, and multiple awards to support observations with HST through the Space Telescope Science Institute.  The resources of the NASA-IPAC Extragalactic Database and HyperLEDA hosted in Lyon, France have been indispensable. 
B.E.S. was supported by the Marc J, Staley graduate fellowship at UC Berkeley.

\section*{Data Availability}

We presented all data underlying this article in three tables in the Appendix. The complete versions of these tables are available within the public domain of the Extragalactic Distance Database (EDD: \url{https://edd.ifa.hawaii.edu/dfirst.php}) where Table~\ref{tab:individual_galaxies} is called All CF4 Individual Distances, Table~\ref{tab:galaxy_groups} is called All CF4 Groups, and Table~\ref{tab:group_vpec} is called CF4 All Group Velocities.


\section{Appendix}
\label{sec:appendix}

An overview of the {\it Cosmicflows-4} products was given in \S \ref{sec:catalog}.  The three tables provide information on (1) each individual galaxy with a distance measurement, (2) the distance properties averaged over members of groups, and (3) inferred properties from the velocities of the groups.

\bigskip
The information on individual galaxies in Table~\ref{tab:individual_galaxies} by column:

1. PGC: The Principal Galaxies Catalog identification of the galaxy in HyperLEDA (https://leda.univ-lyon1.fr).

2. 1PGC: The PGC ID of the dominant galaxy in the group containing the galaxy in question given directly or inferred from the group catalogs of \citet{2017ApJ...843...16K} or \citet{2015AJ....149..171T}.

3. T17: The group identification given by \citet{2017A&A...602A.100T}.

4. $V_{cmb}$: Systemic velocity of the galaxy in the reference frame of the cosmic microwave background.

5. DM: Distance modulus of the galaxy derived from an MCMC analysis incorporating all methodologies and associated uncertainty.

6$-$13. DM$_{xxx}$: Distance moduli and uncertainties of the following methodologies after registration to a common scale with the MCMC analysis: (6) SN~Ia, (7) TF, (8) FP, (9) SBF, (10) SN~II, (11) TRGB, (12) CPLR, (13) MASER.

14$-$19. Celestial, galactic, and supergalactic coordinates of the galaxy.

\bigskip
The information on groups of galaxies in Table~\ref{tab:galaxy_groups} by column:

1. 1PGC: The PGC identification of the dominant galaxy in the group in question, given directly or inferred from the group catalogs of \citet{2017ApJ...843...16K}, \citet{2015AJ....149..171T}, and \citet{2017A&A...602A.100T}.  If from the latter source, the 1PGC identification is associated with the brightest galaxy in the {\it Cosmicflows-4} collection.

2. DM: The distance modulus and uncertainty of the group, weighted averaged over all members of the group with measured distance moduli.

3. $V_{cmb}$: Systemic velocity of group in frame of the CMB.

4$-$16. Number of galaxies in group with a distance contribution by a specific methodology and the weighted average modulus of the specific modulus, with uncertainties.  Columns by methodology are (4),(5) TRGB, CPLR, or MASER calibrator, (6),(7) SNIa, (8),(9) FP, (10),(11) TF, (12-13) SBF from optical observations, (14 and 15) SBF from infrared observations, (16) SN~II (never more than one occurrence in these groups). 

\bigskip
The information on velocity-related information pertaining to the groups in Table~\ref{tab:group_vpec} by column:

1. 1PGC: The PGC identification of the dominant galaxy in group as in Table~\ref{tab:galaxy_groups}, Column 1.

2. DM: The same distance modulus and uncertainty of the group averaged over all members of the group with measured distance moduli as in Table~\ref{tab:galaxy_groups}, Column 2.

3. $d$: The luminosity distance corresponding to the modulus in Column 2.

4$-$7. $V_{xxx}$: Group velocities in respectively the reference frames of the Sun, the Local Sheet \citep{2008ApJ...676..184T}, and the cosmic microwave background. The value in column 7 is the value in column 6 multiplied by the cosmological curvature adjustment parameter $f_i = 1+1/2[1-q_{0}]z_{i}-1/6[1-q_{0}-3q^{2}_{0}+j_{0}]z_{i}^{2}$, where $z_{i}$ is the redshift of the galaxy, $q_{0}$ and $j_{0}$ are the acceleration and jerk parameters and $c$ is the speed of light.  Values for the matter and vacuum energy density of the universe $\Omega_m=0.27$ and $\Omega_{\Lambda}=0.73$ are assumed.

8$-$10. $V_{pxx}$: The peculiar velocity of the group following from Eqs. \ref{eq:vds}, \ref{eq:vwf}, and \ref{eq:vramp}, respectively.  The latter is the reference peculiar velocity of this study.

11. $H_i$: Group Hubble parameter, $H_i = f_iV_{cmb,i}/d_i$.

12. Log $H_i$: Logarithm of group Hubble parameter.

13$-$18. Celestial, galactic, and supergalactic coordinates of group.

19$-$21. SGX, SGY, SGZ: Cartesian supergalactic coordinates of group in observed velocity units.

\begin{turnpage}
\setlength{\tabcolsep}{0.1cm}
\tabletypesize{\scriptsize}


\begin{deluxetable*}{rrrrcccccccccrrrrrr}
\centering
\scriptsize
\tablecaption{Distances of Individual Galaxies$^{\dag}$}
\label{tab:individual_galaxies}
\tablehead{ \\
PGC &
1PGC &
T17 &
V$_{cmb}$ &
DM &
DM$_{snIa}$ &
DM$_{tf}$ &
DM$_{fp}$ &
DM$_{sbf}$ &
DM$_{snII}$ &
DM$_{trgb}$ &
DM$_{cplr}$ &
DM$_{maser}$ &
RA &
DE &
glon &
glat &
sgl &
sgb 
\\
 &
 &
 &
(km/s) &
(mag) &
(mag) &
(mag) &
(mag) &
(mag) &
(mag) &
(mag) &
(mag) &
(mag) &
(deg) &
(deg) &
(deg) &
(deg) &
(deg) &
(deg) 
\\
(1) & (2) & (3) & (4) & (5) & (6) & (7) & (8) & (9) & (10) & (11) & (12) & (13) & (14) & (15) & (16) & (17) & (18) & (19)
}
\startdata
       2 &   73150 &      0 &  4726 & 34.535$\pm$0.460 &               & 34.535$\pm$0.46 &               &                &               &              &                &              &   0.0070 &  47.2745 & 113.9553 & -14.6992 & 341.6440 &  20.7388 \\
      4 &     120 &      0 &  4109 & 33.495$\pm$0.390 &               & 33.495$\pm$0.39 &               &                &               &              &                &              &   0.0144 &  23.0876 & 107.8322 & -38.2729 & 316.0587 &  18.4514 \\
     12 &      12 &      0 &  6195 & 34.995$\pm$0.410 &               & 34.995$\pm$0.41 &               &                &               &              &                &              &   0.0358 &  -6.3739 &  90.1920 & -65.9300 & 286.4249 &  11.3511 \\
     16 &      16 &      0 &  5312 & 34.655$\pm$0.450 &               & 34.655$\pm$0.45 &               &                &               &              &                &              &   0.0471 &  -5.1587 &  91.6006 & -64.8655 & 287.6120 &  11.7030 \\
     35 &      35 &      0 &    34 & 29.730$\pm$0.100 &               &               &               &                &               & 29.73$\pm$0.10 &                &              &   0.0938 &  39.4955 & 112.3102 & -22.3207 & 333.3464 &  20.3429 \\
     55 &      55 &      0 &  4454 & 34.245$\pm$0.390 &               & 34.245$\pm$0.39 &               &                &               &              &                &              &   0.1558 &  33.6009 & 110.9496 & -28.0857 & 327.0996 &  19.7763 \\
     64 &   72642 &      0 & 15324 & 36.606$\pm$0.630 &               &               & 36.606$\pm$0.63 &                &               &              &                &              &   0.2180 & -35.8436 & 350.7982 & -76.1593 & 258.4801 &   1.3810 \\
     66 &   72642 &      0 & 14725 & 36.276$\pm$0.590 &               &               & 36.276$\pm$0.59 &                &               &              &                &              &   0.2215 & -35.9863 & 350.3084 & -76.0782 & 258.3474 &   1.3282 \\
     68 &      68 &      0 &  7338 & 34.735$\pm$0.560 &               & 34.735$\pm$0.56 &               &                &               &              &                &              &   0.2306 & -18.9589 &  65.4189 & -75.8101 & 274.3903 &   7.1770 \\
     70 &      70 &      0 &  6447 & 35.325$\pm$0.390 &               & 35.325$\pm$0.39 &               &                &               &              &                &              &   0.2336 &  20.3380 & 107.1780 & -40.9837 & 313.2487 &  17.7662 \\
\nodata \\
\enddata
\tablenotetext{$\dag$}{The complete version of this table is available online and also as the catalog All CF4 Individual Distances within the Extragalactic Distance Database (\url{https://edd.ifa.hawaii.edu}).
The following catalogs are available pertaining to the separate methodologies: column 6 All CF4 SN~Ia Samples, column 7 All CF4 TFR Samples, column 8 All CF4 FP Samples, column 9 All CF4 SBF Samples (optical) and HST IR SBF (infrared), column 10 de Jaeger SN~II, column 11 All CF4 TRGB. column 12 All CF4 Cepheid, and column 13 All CF4 Maser. Distances in the catalogs for separate methodologies are on arbitrary zero-point scales.
}
\end{deluxetable*}


\begin{deluxetable*}{rcrrcrcrcrcrcrcc}
\scriptsize
\tablecaption{Distances of Galaxy Groups$^{\dag}$}
\label{tab:galaxy_groups}
\tablehead{ \\
1PGC &
DM &
V$_{cmb}$ &
N$_c$ &
DM$_{cal}$ &
N$_{snIa}$ &
DM$_{snIa}$ &
N$_{fp}$ &
DM$_{fp}$ &
N$_{tf}$ &
DM$_{tf}$ &
N$_{sbfo}$ &
DM$_{sbfo}$ &
N$_{sbfi}$ &
DM$_{sbfi}$ &
DM$_{snII}$ 
\\
 & mag & km/s & & mag & & mag & & mag & & mag & & mag & & mag & mag \\
(1) & (2) & (3) & (4) & (5) & (6) & (7) & (8) & (9) & (10) & (11) & (12) & (13) & (14) & (15) & (16)
}
\startdata
  44715 &  34.891$\pm$0.025 &   7193 &    0 &   0.000 &    7 &  34.96$\pm$0.06 &  209 &  34.964$\pm$0.033 &   50 &  34.79$\pm$0.07 &    0 &    &   2 &  34.865$\pm$0.077 &    \\
  56962 &  36.004$\pm$0.028 &  11266 &    0 &   0.000 &    2 &  36.05$\pm$0.11 &  193 &  36.065$\pm$0.035 &   60 &  35.91$\pm$0.05 &    0 &    &   0 &                 &    \\
  58265 &  35.498$\pm$0.038 &   9197 &    0 &   0.000 &    2 &  35.68$\pm$0.12 &  146 &  35.510$\pm$0.040 &    2 &  35.72$\pm$0.29 &    0 &    &   0 &                 &    \\
  36071 &  36.980$\pm$0.042 &  17476 &    0 &   0.000 &    1 &  37.02$\pm$0.18 &  129 &  37.006$\pm$0.043 &    0 &               &    0 &    &   0 &                 &    \\
  55151 &  37.373$\pm$0.046 &  19859 &    0 &   0.000 &    0 &               &  115 &  37.393$\pm$0.046 &    0 &               &    0 &    &   0 &                 &    \\
  47202 &  36.392$\pm$0.051 &  14695 &    0 &   0.000 &    1 &  36.84$\pm$0.13 &   93 &  36.391$\pm$0.064 &   13 &  36.22$\pm$0.11 &    0 &    &   0 &                 &    \\
  45753 &  37.423$\pm$0.057 &  21182 &    0 &   0.000 &    0 &               &   77 &  37.443$\pm$0.057 &    0 &               &    0 &    &   0 &                 &    \\
  51335 &  37.395$\pm$0.056 &  21561 &    0 &   0.000 &    0 &               &   74 &  37.415$\pm$0.056 &    0 &               &    0 &    &   0 &                 &    \\
  47982 &  37.220$\pm$0.056 &  21615 &    0 &   0.000 &    0 &               &   73 &  37.240$\pm$0.056 &    0 &               &    0 &    &   0 &                 &    \\
  42543 &  37.416$\pm$0.057 &  21345 &    0 &   0.000 &    0 &               &   71 &  37.436$\pm$0.057 &    0 &               &    0 &    &   0 &                 &    \\
\nodata \\
\enddata
\tablenotetext{$\dag$}{The complete version of this table is available online and also as the catalog CF4 All Groups within the Extragalactic Distance Database (\url{https://edd.ifa.hawaii.edu}).  The following catalogs are available pertaining to the separate methodologies: column 5 CF4 All Calibrator Groups, column 7 CF4 SN~Ia Groups, column 9 CF4 FP Groups, column 11 CF4 TFR Groups, columns 13 and 15 CF4 SBF Groups, column 16 CF4 SN~II Group". Distances in the catalogs for separate methodologies are on arbitrary zero-point scales.
}
\end{deluxetable*}


\begin{deluxetable*}{rcrrrrrrrrrcrrrrrrrrr}
\centering
\scriptsize
\tablecaption{Galaxy Group Distances and Peculiar Velocities$^{\dag}$}
\label{tab:group_vpec}
\tablehead{ \\
1PGC &
DM &
d &
V$_{hel}$ &
V$_{ls}$ &
V$_{cmb}$ &
fV &
V$_{pds}$ &
V$_{pwf}$ &
V$_{pec}$ &
H$_i$ &
log H &
RA &
DE &
glon &
glat &
sgl &
sgb &
SGX &
SGY &
SGZ
\\
 &
mag &
Mpc &
km/s &
km/s &
km/s &
km/s &
km/s &
km/s &
km/s &
km/s/Mpc &
km/s/Mpc &
deg &
deg &
deg &
deg &
deg &
deg &
Mpc &
Mpc &
Mpc 
\\
(1) & (2) & (3) & (4) & (5) & (6) & (7) & (8) & (9) & (10) & (11) & (12) & (13) & (14) & (15) & (16) & (17) & (18) & (19) & (20) & (21) }
\startdata
     12 & 34.995$\pm$0.410 &  99.8 &  6532 &  6669 &  6179 &  6280 &  -1174 &  -468 &  -468 &  62.9 & 1.799 &   0.0360 &  -6.3739 &  90.1922 & -65.9300 & 286.4249 &  11.3510 &   1848 &  -6271 &   1312 \\
     14 & 36.095$\pm$0.380 & 165.6 & 11837 & 12035 & 11474 & 11820 &   -575 &  -244 &  -244 &  71.4 & 1.854 & 359.9805 &   8.1846 & 101.6600 & -52.5465 & 300.8070 &  15.3909 &   5942 &  -9967 &   3194 \\
     16 & 34.655$\pm$0.450 &  85.3 &  5709 &  5851 &  5354 &  5430 &   -948 &  -380 &  -380 &  63.6 & 1.804 &   0.0470 &  -5.1587 &  91.6006 & -64.8655 & 287.6120 &  11.7030 &   1733 &  -5461 &   1187 \\
     35 & 29.730$\pm$0.100 &   8.8 &   335 &   618 &    28 &    28 &   -633 &   -38 &  -510 &   3.2 & 0.501 &   0.0938 &  39.4955 & 112.3102 & -22.3207 & 333.3464 &  20.3429 &    518 &   -260 &    215 \\
     55 & 34.245$\pm$0.390 &  70.6 &  4791 &  5064 &  4466 &  4519 &   -765 &  -307 &  -307 &  64.0 & 1.806 &   0.1558 &  33.6010 & 110.9495 & -28.0857 & 327.0996 &  19.7763 &   4001 &  -2589 &   1713 \\
     63 & 35.245$\pm$0.410 & 111.9 &  8743 &  9004 &  8404 &  8590 &    189 &    83 &    83 &  76.7 & 1.885 & 359.9938 &  28.2883 & 109.3762 & -33.2252 & 321.4869 &  19.2652 &   6651 &  -5294 &   2971 \\
     68 & 34.735$\pm$0.560 &  88.5 &  7684 &  7759 &  7358 &  7501 &    844 &   388 &   388 &  84.7 & 1.928 &   0.2305 & -18.9590 &  65.4189 & -75.8101 & 274.3903 &   7.1770 &    589 &  -7676 &    969 \\
     70 & 35.325$\pm$0.390 & 116.1 &  6803 &  7043 &  6450 &  6560 &  -2090 &  -791 &  -791 &  56.5 & 1.752 &   0.2334 &  20.3381 & 107.1779 & -40.9836 & 313.2488 &  17.7663 &   4595 &  -4886 &   2149 \\
     75 & 36.384$\pm$0.345 & 189.1 & 11750 & 11700 & 11521 & 11870 &  -2212 &  -884 &  -884 &  62.8 & 1.798 &   0.3388 & -43.6117 & 331.2258 & -70.6612 & 251.2392 &  -1.4353 &  -3762 & -11074 &   -293 \\
     76 & 34.885$\pm$0.390 &  94.8 &  6903 &  7166 &  6566 &  6680 &   -423 &  -178 &  -178 &  70.4 & 1.848 &   0.2454 &  28.9118 & 109.8059 & -32.6706 & 322.1729 &  19.1316 &   5347 &  -4152 &   2348 \\
\nodata \\
\enddata
\tablenotetext{${\dag}$}{The complete version of this table is available online and also as the catalog CF4 All Group Velocities within the Extragalactic Distance Database (\url{https://edd.ifa.hawaii.edu}).}
\end{deluxetable*}

\end{turnpage}

\clearpage

\bibliography{main}
\bibliographystyle{aasjournal}

\end{document}